\documentclass[useAMS,usenatbib]{mnras}
\usepackage{graphicx}
\usepackage{amsmath}
\usepackage{amssymb}

\def\gsim{\;\rlap{\lower 2.5pt
 \hbox{$\sim$}}\raise 1.5pt\hbox{$>$}\;}
\def\lsim{\;\rlap{\lower 2.5pt
   \hbox{$\sim$}}\raise 1.5pt\hbox{$<$}\;}
   
\newcommand{\tr}[1]{\textrm{#1}}
\newcommand{\ee}[1]{\times10^{#1}}

\newcommand{\pp}[2]{\frac{\partial#1}{\partial#2}}

\title{Cosmic Ray Acceleration of Cool Clouds in the Circumgalactic Medium}
\author[Wiener, Zweibel, \& Ruszkowski]{Joshua Wiener$^{1}$, Ellen G. Zweibel$^{1,2}$ \& Mateusz Ruszkowski$^3$\\
$^{1}$ Department of Astronomy, University of Wisconsin-Madison, Madison, WI 53706, USA.\\
$^{2}$ Department of Physics, University of Wisconsin-Madison, Madison, WI 53706, USA.\\
$^3$ Department of Astronomy, University of Michigan, Ann Arbor, MI 48109, USA.}

\begin{document}

\maketitle

\defcitealias{wiener17a}{W17}

\begin{abstract}
We investigate a mechanism for accelerating cool (10$^4$ K) clouds in the circumgalactic medium (CGM) with cosmic rays (CRs), possibly explaining some characteristics of observed high velocity clouds (HVCs). Enforcing CRs to stream down their pressure gradient into a region of slow streaming speed results in significant buildup of CR pressure which can accelerate the CGM. We present the results of the first two-dimensional magnetohydrodynamic (MHD) simulations of such `CR bottlenecks,' expanding on simpler simulations in 1D from \cite{wiener17a}. Although much more investigation is required, we find two main results. First, radiative cooling in the interfaces of these clouds is sufficient to keep the cloud intact to CR wave heating. Second, cloud acceleration depends almost linearly with the injected CR flux at low values (comparable to that expected from a Milky Way-like star formation rate), but scales sublinearly at higher CR fluxes in 1D simulations. 2D simulations show hints of sublinear dependence at high CR fluxes but are consistent with pure linear dependence up to the CR fluxes tested. It is therefore plausible to accelerate cool clouds in the CGM to speeds of hundreds of km s$^{-1}$.
\end{abstract}

\section{Introduction}
The circumgalactic medium (CGM), a halo of gas surrounding the Galactic disk, is an important component of any model of galaxy evolution. The CGM contains most of the baryons in the Galaxy and serves as a potential reservoir of fresh material to form new stars. At the same time, the chemical makeup of the CGM holds imprints of the Galaxy's star formation history, as supernova eject material into the CGM in large-scale Galactic fountains or outflows. The makeup and structure of the CGM is thus of great interest \citep{bregman18,zhang18}.

One of many probes into the CGM is absorption line spectroscopy of bright background (extraglactic) sources. \cite{wakker12} analyze dozens of sight lines through the Galactic halo and use absorption line features to measure the column densities of various ions and their velocity structure. Among other findings, they found large column densities of OVI, suggesting the existence of ``transition temperature'' gas at 10$^{5.5}$ K. Additionally, Doppler shifts in the absorption lines suggest bulk velocities of up to 100 km s$^{-1}$ for some components. The multiphase nature of the CGM of other galaxies is revealed in observations of sight-lines through galaxy halos, such as the COS-Halos survey. \cite{werk14} examine low-ionization state absorption lines to characterize the cool 10$^4$ K component of the CGM and find that the electron density of this cool phase is 100 times lower than what is expected from thermal equilibrium with the hot 10$^6$ K phase. \cite{werk17} study the high-ionization state component using OVI absorption lines and find that while many OVI absorption features are cospatial with low-ionization absorption, some are not, suggesting the presence of a 10$^{5.5}$ K phase. Low-ionization features are also observed in galactic outflows (see \cite{chisholm16}), suggesting that galactic winds also contain multiple gas phases.

From these and other observations, astronomers infer the presence of cool or warm (10$^4$ K) clouds moving very quickly within the otherwise hot (10$^6$ K) CGM, with thick intermediate temperature (10$^{5.5}$ K) interfaces. These are referred to as high velocity clouds (HVCs). The properties of HVCs introduce some puzzling questions about the dynamics of the CGM. Were they accelerated along with the launching of a Galactic wind, or did they condense \emph{in situ} out of the hot phase of a previously launched wind? We consider here only the former possibility, addressing the issue that Galactic wind models using thermal pressure alone have trouble accelerating cool clouds to the 100 km s$^{-1}$ speeds measured without destroying them (for example, see \cite{scannapieco15}). Additionally, few models are able to generate high OVI column densities, or even the right ion ratios \citep{wakker12}.

In \cite{wiener17a} (hereafter W17) we investigated the influence of cosmic ray (CR) transport physics in this environment. Using simple 1D models, we showed that due to a bottleneck effect arising from the spatially-varying CR streaming speed, large CR pressure gradients build up as CRs exit a cool cloud. This gradient has dynamic effects, accelerating the cloud to high velocities consistent with observations, as well as energetic effects, heating the interface at the location of the gradient in such a way as to nearly reproduce observed ion ratios. 

These bottlenecks may also significantly impact the spatial distribution of CR pressure in a galactic wind, critically affecting the nature of CR-driven winds. CR-driven winds are advantageous over thermally-driven winds since CRs are less subject to radiative losses and maintain their energy over longer distances. Generally speaking, CR-driven wind models involve weak CR pressure gradients pushing on the gas gently over these long distances, resulting in wind velocities that increase with height above the disk. However, as we found in \citetalias{wiener17a}, CR bottlenecks redistribute the CR pressure in space, resulting in sharp CR gradients at one edge of the cloud, and eliminating CR gradients upwind of this point. As such the nature of a CR driven wind through a medium with many such bottlenecks could be radically different from one without them.

In this work we expand upon these simple 1D hydrodynamic (HD) simulations with a set of 2D magnetohydrodynamic (MHD) simulations using the {\small FLASH} code, modified to include CR streaming physics (\cite{ruszkowski17}). Unlike in \citetalias{wiener17a}, these simulations handle radiative cooling in a self-consistent way, and account for the potential bending of magnetic field lines as the system evolves. This is important in regards to the bottleneck effect, since the amount of magnetic flux penetrating the cloud controls how many CRs go into the bottleneck. A magnetically-isolated cloud, for example, would have no CR bottleneck at all. By performing simulations in 2D (and possibly in future work, 3D) we can examine the impact of these lateral effects.

In \S\ref{sec:theory} we briefly review the CR streaming physics that result in the bottleneck effect. In \S\ref{sec:setup} we describe the {\small FLASH} code and the simulation setups. In \S\ref{sec:results} we present the results of our simulations and discuss their implications, focusing on the difference with our simple 1D simulations. We conclude in \S\ref{sec:conclusion}. Effects of our boundary conditions and resolution tests are presented in Appendices \ref{sec:boundary} and \ref{sec:resolution} respectively.

\section{Cosmic Ray Streaming}\label{sec:theory}
All the scales in this problem are much larger than the scattering mean free path of individual CRs, so we will describe the CRs in a simple fluid approximation. The CRs are treated as a single fluid with pressure $P_c$ and adiabatic index $\gamma_c$, assumed to be 4/3. We evolve this pressure in time along with a background gas described by density $\rho$, velocity $u$, and internal energy $E_g$, as well as a magnetic field $B$.

We use the self-confinement theory of CR transport, wherein the bulk flow of CRs is regulated by the streaming instability (\cite{kulsrud69,skilling71,kulsrud05,zweibel17}). In this framework, Alfv\'en waves are amplified by gyroresonant cosmic rays wherever the bulk CR flow speed along the magnetic field $v_s$ exceeds the local Alfv\'en speed $v_A$. These waves then scatter CRs in pitch angle until $v_s$ falls below $v_A$ and the instability shuts off. Moreover, in the limit of short mean free path, the streaming speed is proportional to the CR pressure gradient. The CR streaming speed $v_s$ is thus determined by the marginal stability condition for the streaming instability, which, in the case where Alfv\'en wave damping processes are negligible, is simply described by $v_s=v_A$ with the constraint that the direction of net CR flow be along the local magnetic field, and down the CR pressure gradient. We then say that CRs are `locked to the wave frame' wherever there is a non-zero CR pressure gradient along the local magnetic field.

One consequence of the self-confinement picture is an energy exchange between CRs and the thermal plasma, mediated by the Alfv\'en waves. One can show that streaming CRs will heat the gas at a rate equal to
\begin{equation}\label{eq:crheat}
\mathcal{H}_c=-\mathbf{v_A}\cdot\nabla P_c=v_A|\mathbf{\hat{b}}\cdot\nabla P_c|.
\end{equation}
This was first done by \cite{wentzel71}. Here, $\mathbf{\hat{b}}$ is a unit vector pointing along the magnetic field. This heating has potential consequences on the energetics of a system, especially at the locations of sharp CR pressure gradients. Such gradients may develop because of a second consequence of the self-confinement picture, the CR bottleneck. For a detailed description of the CR bottleneck effect, which was first predicted by \cite{skilling71}, we refer the reader to \citetalias{wiener17a}, but we summarize the main points here. Consider a one-dimensional flow of CRs along a single magnetic flux tube. As $B$ or $\rho$ change along the tube, the streaming speed $v_A$ changes. If $v_A$ increases along the direction of the CR flow, the CR pressure $P_c$ must drop according to
\begin{equation}\label{eq:cr1D}
\mathbf{\hat{b}}\cdot\mathbf{\nabla}P_c\rho^{-\gamma_c/2}=0.
\end{equation}
If $\rho$ increases along the direction of the CR flow, we might expect $P_c$ to increase accordingly. But this would violate the constraint that CRs only stream down their gradient. Therefore, the only steady state solution in this case is
\begin{equation}
P_c=\tr{constant}.
\end{equation}
A maximum in $\rho$ along a flux tube therefore produces a bottleneck. In the steady state, $P_c$ `upstream' of the maximum must be constant. `Downstream' of the maximum $P_c$ follows \eqref{eq:cr1D}. This effect will be present in cool clouds in the CGM provided the magnetic field lines penetrate the cloud. We would then expect $\rho$ to have a maximum in the center of the cloud, causing such a CR bottleneck, and building up a strong pressure gradient on one end of the cloud. 

This picture is subject to a number of caveats. We do not include any wave damping in our analysis or simulations, but  inside the cloud it may be cold enough for Alfv\'en waves to suffer significant ion-neutral damping. At late times we expect the bottleneck to erase CR pressure gradients in this region (we will see this in the simulations in \S\ref{sec:results}). Waves are not excited in the first place, so the presence or absence of ion-neutral damping has no effect on the dynamics at this point. However, at early times ion-neutral damping may have a transient effect by allowing CRs to ``fill up'' the cloud region more rapidly. More importantly, bottlenecks can be completely erased if magnetic field lines do not penetrate the cloud at all. The field line topology of a cool cloud will depend strongly on how it is formed. We refer the reader to \citetalias{wiener17a} for further discussion of these issues.

\section{Simulation Setup}\label{sec:setup}
\subsection{Basic Equations}\label{subsec:basics}
We use the magnetohydrodynamics (MHD) code {\small FLASH} in 2D to solve the ideal MHD equations with modifications to include CRs and CR streaming, according to \cite{ruszkowski17} as well as thermal gas heating and cooling terms. The equations we solve are as in \cite{ruszkowski17}, but without gravity ($\mathbf{g}=0$), explicit CR diffusion ($\mathbf{\kappa}=0$), or supernova feedback ($\dot{p}_\tr{SN}=\mathcal{H}_\tr{SN}=0$):
\begin{equation}
\pp{\rho}{t}+\nabla\cdot(\rho\mathbf{u})=0
\end{equation}
\begin{equation}
\pp{\rho\mathbf{u}}{t}+\nabla\cdot\left(\rho\mathbf{uu}-\frac{\mathbf{BB}}{4\pi}\right)+\nabla P_\tr{tot}=0
\end{equation}
\begin{equation}
\pp{\mathbf{B}}{t}-\nabla\times(\mathbf{u}\times\mathbf{B})=0
\end{equation}
\begin{equation}
\pp{E}{t}+\nabla\cdot\left[(E+P_\tr{tot})\mathbf{u}-\frac{\mathbf{B}(\mathbf{B}\cdot\mathbf{u})}{4\pi}\right]=-\nabla\cdot\mathbf{F}_c-\mathcal{C}
\end{equation}
\begin{equation}
\pp{E_c}{t}+\nabla\cdot(E_c\mathbf{u})=-P_c\nabla\cdot\mathbf{u}-\mathcal{H}_c-\nabla\cdot\mathbf{F}_c+q_c
\end{equation}
In the above $\rho$ and $\mathbf{u}$ are the gas mass density and velocity, $\mathbf{B}$ is the magnetic field, and $P_\tr{tot}$ is the total pressure
\begin{equation}
P_\tr{tot}=P_g+P_c+\frac{B^2}{8\pi}
\end{equation}
consisting of thermal pressure $P_g=\rho k_B T/(\mu m_p)$, CR pressure $P_c$, and magnetic pressure. The total energy density $E$ is
\begin{equation}
E=E_g+E_c+\frac{1}{2}\rho u^2+\frac{B^2}{8\pi}
\end{equation}
and includes the thermal energy density $E_g=P_g/(\gamma_g-1)$, the CR energy density $E_c=P_c/(\gamma_c-1)$, the kinetic energy density and the magnetic energy density. The CR flux $\mathbf{F}_c$ is written in our model as a simple advective flux
\begin{equation}
\mathbf{F}_c=(E_c+P_c)\mathbf{v}_s
\end{equation}
where the streaming speed $\mathbf{v}_s$ has magnitude up to $v_A$ (see below) and points along the local magnetic field. A CR source term $q_c$ is included, which we use to add CRs to the domain (see \S\ref{subsec:specifics}). Lastly, $\mathcal{C}$ represents radiative cooling and a heating function which maintains thermal equilibrium in the absence of cosmic rays, discussed below, and $\mathcal{H}_c$ is CR heating, given by \eqref{eq:crheat}.

As a brief aside, let us elaborate on the CR streaming velocity $\mathbf{v_s}$, both in the self-confinement model and in the {\small FLASH} code implementation. In the model, which neglects wave damping, CRs will stream along the background magnetic field at the Alfv\'en speed $v_A$ anywhere there is a CR pressure gradient along the field to drive the flow. If the CR pressure is constant along the field, $\mathbf{v_s}$ will still be along the field, but will have magnitude between 0 and $v_A$, determined by other factors such as a steady-state requirement. For instance, as discussed in \S\ref{sec:theory}, the steady state solution upstream of a bottleneck is $P_c=$constant, $|v_s|=v_{A,\tr{min}}<v_{A,\tr{local}}$.

In simulation, the requirement that CRs only flow down their gradient presents a computational challenge which in the CR module in {\small FLASH} (see \cite{ruszkowski17}) is handled with a regularization method described in \cite{sharma10}. In this method a scale length $L$ is chosen, and the streaming speed is determined by
\begin{equation}\label{eq:tanh}
v_s=v_A\tanh \left(\frac{|\mathbf{\hat{b}}\cdot\nabla P_c|}{P_c/L}\right)
\end{equation}
The length scale sets a characteristic CR pressure gradient $P_c/L$ - gradients much steeper than this will have $v_s=v_A$, while smaller gradients will have $v_s<v_A$. In this way the streaming speed is consistently calculated everywhere, and it exhibits the behavior required by the self-confinement model: upstream of a bottleneck, CR pressure builds up, approaching a flat profile. The reduced CR gradient in this region results in the streaming speed $v_s$ dropping according to \eqref{eq:tanh} above. This emulates the theoretical steady state picture in an approximate way which depends on the chosen length scale $L$. The larger $L$ is, the more accurate the simulation, but the more restrictive the time step. There are other ways to treat CR streaming - \cite{jiang18} introduce a two-moment method where the CR flux $\mathbf{F}_c$ is treated as an independent variable with its own evolution equation. We briefly discuss this in \S\ref{sec:future}.

\subsection{Specifics of Model}\label{subsec:specifics}
The background plasma is initialized at uniform pressure $P_g=3.23\ee{-13}$ dyne cm$^{-2}$ and mass density $\rho_\tr{hot}=2.1\ee{-27}$ g cm$^{-3}$. We assume hydrogen and helium are fully ionized. We use an average mass per particle of $\mu=0.60m_p$, and average mass per Hydrogen nucleus of $\mu_H=1.43m_p$. This mass density and thermal pressure therefore correspond to a Hydrogen number density of $n_H=9.0\ee{-4}$ cm$^{-3}$ and a temperature $T=1.1\ee{6}$ K. In this background we insert a circular, fully-ionized,  cooler cloud at the same pressure but with mass density $\rho_\tr{cold}=2.35\ee{-25}$ g cm$^{-3}$, corresponding to a density $n_H=0.10$ cm$^{-3}$ and a temperature $T=9.9\ee{3}$ K\footnote{Please note that while we quote three significant figures for some quantities in the spirit of transparency, the derived quantities are only held to two significant figures.}. The density follows a tanh profile of radius $r_c=50$ pc and interface thickness $t_c=0.25$ pc according to
\begin{equation}\label{eq:pulse}
\rho(r)=\rho_\tr{hot}+(\rho_\tr{cold}-\rho_\tr{hot})\left(\frac{1}{2}-\frac{1}{2}\tanh\left(\frac{r-r_c}{t_c}\right)\right).
\end{equation}
This initial density profile is a 2D generalization of the one used in \citetalias{wiener17a}. The cloud is centered on coordinate $(x,y)=(1\ \tr{kpc},0)$. The magnetic field is initialized to a uniform value of 1 $\mu$G throughout the domain, pointing in the (horizontal) $x$-direction. With these parameters, $v_A=62$ km s$^{-1}$ in the hot phase. The Alfv\'en travel time from the left boundary to the cloud is $\sim 15$ Myr, while the adiabatic acoustic travel time is $\sim 6$ Myr. These timescales are important for understanding the evolution of the system.

The initial CR distribution is set to a very low, uniform pressure. Each time step, CRs are injected into all the leftmost active cells of the domain via the source term $q_c$. The CR flux through the left boundary is forced to zero to ensure that all injected CRs stay in the domain and flow towards the cloud (this also prevents CRs from advecting into the domain with the gas, so that the amount of CR energy  added to the domain is never more than what is added via the source term). The injection is time-dependent, emulating a sudden burst of star formation, according to
\begin{equation}\label{eq:qcr}
q_c(t)=q_0\left(1-e^{-t/t_\tr{ramp}}\right)\left(1-e^{(t-t_\tr{end})/t_\tr{ramp}}\right)
\end{equation}
with $t_\tr{ramp}=3$ Myr and $t_\tr{end}=30$ Myr. At $t$ greater than $t_\tr{end}$, the CR source is set to zero. The duration of the burst exceeds both the acoustic and Alfv\'en travel times, but not by large factors; thus, a steady state is never achieved. The source strength is set to $q_0=5.0\ee{-25}$ erg cm$^{-3}$ s$^{-1}$ for our fiducial run. We increase this by factors of 3 and 10 in comparison runs. The region of CR injection extends from the left boundary at $x=0$ to $x=D=1$ pc corresponding to a CR injection rate per unit area of $q_0D=1.54\ee{-6}$ erg cm$^{-2}$s$^{-1}$ for the fiducial run. In a 1D situation, we expect the cosmic ray pressure between the cloud and the source to build up to a value of order $q_0D/(3v_A)\sim 8.2\times 10^{-14}$ dyne cm$^{-2}$. This is borne out by Fig. \ref{fig:1xseries}, and is only about 25\% of the thermal pressure, suggesting that the there will be
little lateral magnetic field line distortion or spreading, and that the dynamics will remain 1D.

As in \citetalias{wiener17a}, the onset of CR injection produces an initial sound wave which propagates outward and tends to reflect off the right boundary of the domain. To prevent the non-physical reflected wave from interfering with our results we place the right boundary at a very large distance (20 kpc). To save computation time, the extent of the domain in the (vertical) $y$-direction is limited to $\pm$125 pc, and we make use of the mesh refinement in {\small FLASH} to reduce resolution far to the right of the cloud. The mesh is fixed in time, and is forced to its maximum resolution of 1.95 pc in the region $0<x<3$ kpc.

\begin{figure}
\includegraphics[width=0.5\textwidth]{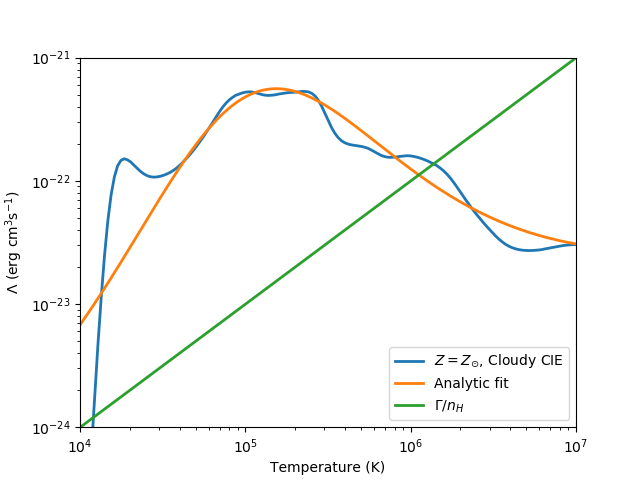}
\caption{Radiative cooling function for solar abundance. The data is from \protect\cite{wiersma09} using {\small CLOUDY}. The fit is a modified version of the analytic function in \protect\cite{bustard16}. Also shown is the supplemental heating, $n_H\Gamma$, divided by $n_H^2$ for the initial thermal pressure used in these simulations, indicating the existence of low temperature and high temperature equilibria.}\label{fig:coolfit}
\end{figure}

\begin{figure*}
\includegraphics[scale=0.5, trim={1cm 0cm 0cm 0cm}]{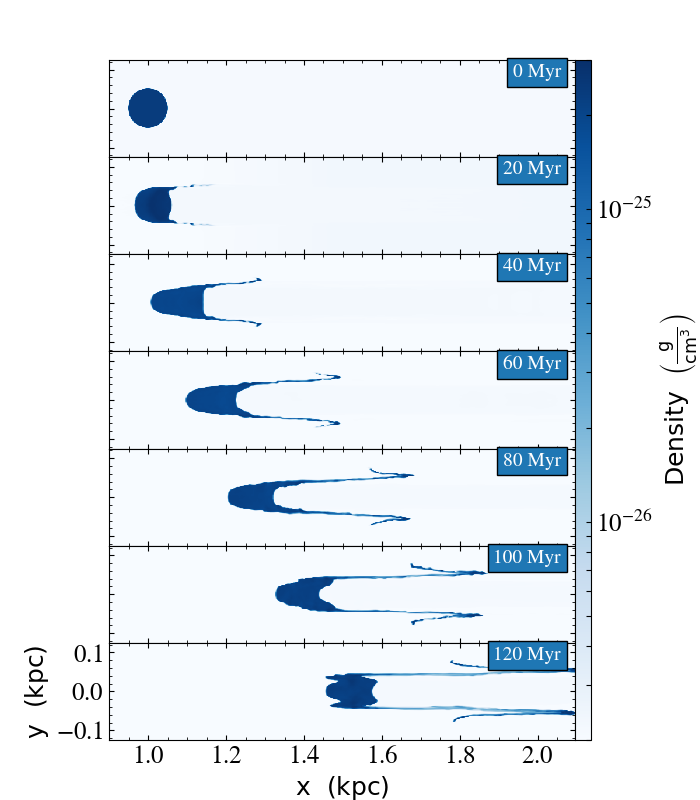}
\includegraphics[scale=0.5, trim={0cm 0cm 1cm 0cm}]{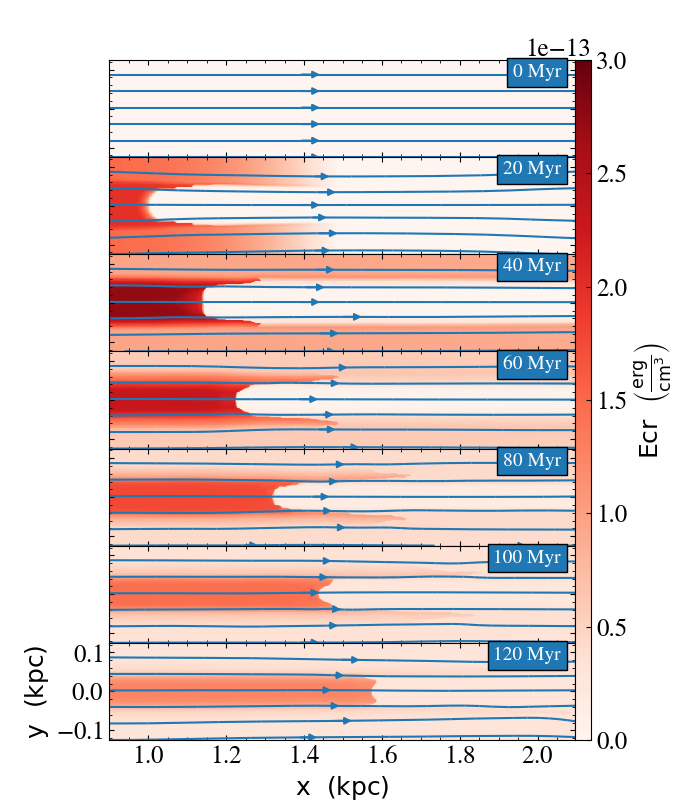}
\caption{Time series slices of the fiducial simulation. A 10$^4$ K cloud in pressure equilibrium with the surrounding 10$^6$ K medium is pushed on by CRs coming in from the left of the domain. CRs are injected at the domain boundary at $x=0$ (not shown) from time $t=0$ to $t=30$ Myr. Left: Mass density $\rho$. Right: CR energy density $E_c=3P_c$. Overlaid are streamlines of the magnetic field.}\label{fig:1xseries}
\end{figure*}

The simulation includes radiative cooling proportional to $n_H^2$, with a supplemental heating rate proportional to $n_H$:
\begin{equation}
\mathcal{C}=n_H^2\Lambda(T)-n_H\Gamma
\end{equation}
The supplemental heating is added to ensure two thermal equilibrium phases at about 10$^4$ and 10$^6$ K. Its form is consistent with the heating expected from photoelectric expulsion of electrons from dust grains. We use a heating constant of $\Gamma=10^{-25}$ erg s$^{-1}$, which is similar to values used in other works (\cite{inoue06} for example). The cooling function is approximated with the analytic function
\begin{equation}
\Lambda(T)=1.1\ee{-21}*10^{\Theta(\log(T/10^5\ \tr{K}))}\tr{ erg cm}^3\tr{ s}^{-1}
\end{equation}
where the exponent is
\[
\Theta(x)=0.4x-3+\frac{5.2}{e^{x+0.08}+e^{-1.5(x+0.08)}}.
\]
This form is taken from \cite{bustard16}, modified by Chad Bustard to fit {\small CLOUDY} data from \cite{wiersma09} (Chad Bustard, private communication). The comparison of this function to the data is shown in figure \ref{fig:coolfit}. We overplot the quantity $\Gamma/n_H$ at the initial thermal pressure given above to show the presence of the two thermal equilibria. We turn off radiative cooling for temperatures below 10$^4$ K. With this setup we have, at the chosen thermal pressure, one thermal equilibrium at exactly 10$^4$ K and one just above 10$^6$ K. The equilibrium at 10$^4$ K is stable. The equilibrium near 10$^6$ K is formally unstable, however the time scales of the growth of deviations from the equilibrium are long compared to the simulation times, so the hot phase of our simulated medium does not change. Although thermal conduction is not included explicitly in our models, numerical diffusion causes the cloud boundary to evolve. This is a small effect on the timescales considered here.

\section{Results}\label{sec:results}
\subsection{Fiducial Run}\label{sec:fid}
The fiducial simulation is run out to 120 Myr, four times the duration of the injected CR pulse. A time series of the density and CR pressure in the vicinity of the cloud is shown in figure \ref{fig:1xseries}. The qualitative picture is the same as the 1D simulations in \citetalias{wiener17a}: the onset of CRs injected at the left run into the cloud, building up a bottleneck. The resulting CR pressure gradient pushes the cloud to the right, reducing the thermal pressure inside until equilibrium is regained.

\begin{figure}
\includegraphics[width=0.5\textwidth, trim=0cm 1cm 0cm 0cm]{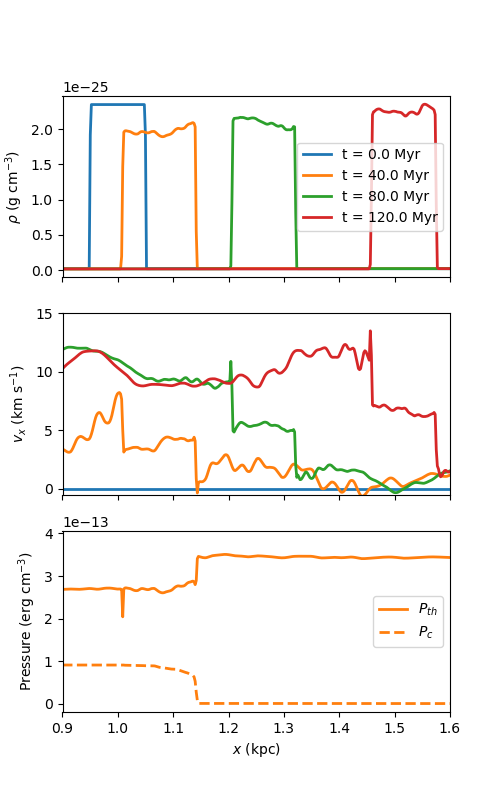}
\caption{Profiles through the center of the cloud ($y=0$) of the fiducial simulation. Top: Mass density $\rho$ at various times. Middle: Horizontal velocity $v_x$ at the same times. Bottom: Thermal pressure $P_{th}$ and CR pressure $P_c$ at $t=40$ Myr, when the bottleneck is at its steepest.}\label{fig:1xvpres}
\end{figure}

The influence of the CRs in this setup turns out to be small. The cloud is only accelerated to speeds of 5-10 km s$^{-1}$, the field lines are only slightly perturbed from their initial configuration, and the CR pressure inside the cloud peaks at about 30\% of the total pressure. This is shown in figure \ref{fig:1xvpres} which plots various quantities along a horizontal slice through the center of the cloud ($y=0$).

The top panel of figure \ref{fig:1xvpres} shows horizontal profiles of the mass density $\rho$ along the line $y=0$, which goes through the center of the cloud. The initial reaction of the cloud to the push from the CRs is to stretch horizontally, reducing the mass density and thermal pressure inside the cloud. This is the same behavior as in the simple 1D simulations in \citetalias{wiener17a}. At later times however the density slowly increases. This may be due to radiative cooling causing condensation onto the cloud.

One of the significant results of this work is the finding that the inclusion of radiative cooling prevents the cloud from being destroyed by CR heating as we saw in the 1D simulations without radiative cooling in \citetalias{wiener17a}, wherein we found that without radiative cooling, CR heating will destroy a cloud in around 100 Myr. This lack of any erosion of the cloud in these simulations suggests it is possible to accelerate clouds to high velocities with this mechanism without destroying them.

The middle panel of figure \ref{fig:1xvpres} shows the horizontal velocity profiles of the gas for different times in the simulation. Other than the initial profile which is uniformly zero, the velocity profiles all show two sharp dropoffs corresponding to the two edges of the cloud. The cloud velocity can be read off as the value of the ``plateau'' between these two edges - it has risen to 4 km s$^{-1}$ by 40 Myr, and accelerates further to about 7 km s$^{-1}$ by the last time step at 120 Myr. These velocities fall far short of the observed 100 km s$^{-1}$ velocities of HVCs (and also the velocities attained in the 1D models). As the bottom panel shows, this fiducial run has a very weak CR source, resulting in the thermal pressure dominating over the CR pressure everywhere. Since this results in very little field line bending, the dynamics are basically one-dimensional. This is evidenced by the qualitative agreement of these profiles with the 1D results from \citetalias{wiener17a}. As to a quantitative comparison, we note that the $\sim 100$ km s$^{-1}$ cloud velocities reported in \citetalias{wiener17a} are reached using a CR source 20 times larger than in this 2D simulation. There are a number of other differences between the 1D and 2D simulations, such as magnetic field and initial cloud position (these factors affect at least the timing of the acceleration), but the difference in source strength is likely the largest factor in the difference in cloud velocities. We discuss this comparison further in \S\ref{sec:higher}.

The bottom panel of figure \ref{fig:1xvpres} shows the thermal and CR pressure profiles at a single time slice at 40 Myr, chosen when the CR bottleneck is around its steepest. As in \citetalias{wiener17a} we can clearly see the bottleneck where the CRs exit the cloud at its right edge, near $x=1.15$ kpc. To the right (downstream) of the bottleneck, CR pressure gradients drive the streaming instability and $P_c$ drops as $v_A$ rises. To the left (upstream) of the bottleneck, the net CR flow is sub-Alfv\'enic, the streaming instability is shut off, and waves are not generated.

Figure \ref{fig:1xvpres} also reveals the limitations of this set of simulations. The 1.95 pc cell size is sufficient to resolve the cloud as a whole, but falls far short of resolving the cloud interfaces. In the interfaces the relevant length scale is set by the balance of radiative cooling with heating mechanisms. Even at the right edge, where the bottleneck provides some CR heating, these length scales are of order a few pc. At the left edge there is no CR pressure gradient and no CR heating, and so the equilibrium length scale is formally zero. Incorporating thermal conduction would alleviate this, but typical Field lengths in this type of system are fractions of a pc at the longest. As Figure 9 of \citetalias{wiener17a} shows, cosmic ray mediated thermal fronts have structure on 0.1 pc scales under the conditions studied here.

The consequences of an underresolved thermal interface are evident in the sharp dip in thermal pressure at the left edge of the cloud shown in the bottom panel of figure \ref{fig:1xvpres}. Because the temperature drops from 10$^6$ K to 10$^4$ K in just a few grid cells, the cooling rate can be very spatially sporadic, resulting in the spiky drops in thermal pressure seen here. Presumably there is a non-physical overabundance of material at temperatures around 10$^{5.5}$ K, at the peak of the cooling curve, resulting in more energy lost to radiative cooling than in a fully resolved simulation. The large scale effects of this are not clear, but we speculate that the strange nature of the swept-up ``wings'' at the top and bottom of the cloud at late times (visible in the left panel of figure \ref{fig:1xseries}) may be due to this.

\subsection{Stronger CR sources}\label{sec:higher}
\begin{figure*}
\includegraphics[scale=0.41]{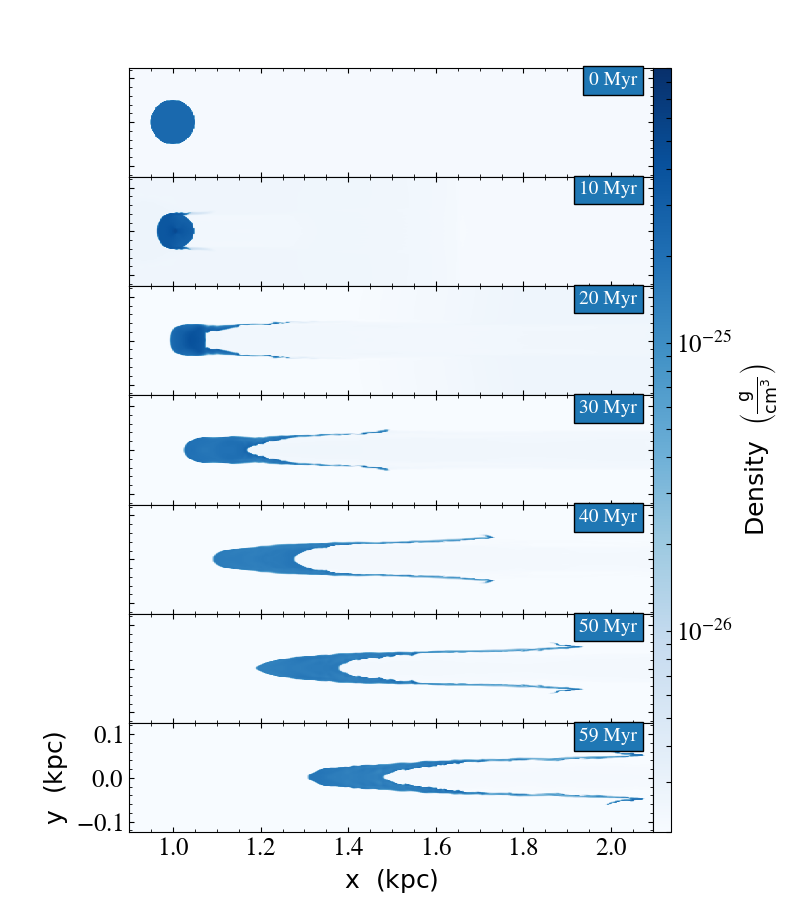}
\includegraphics[scale=0.41]{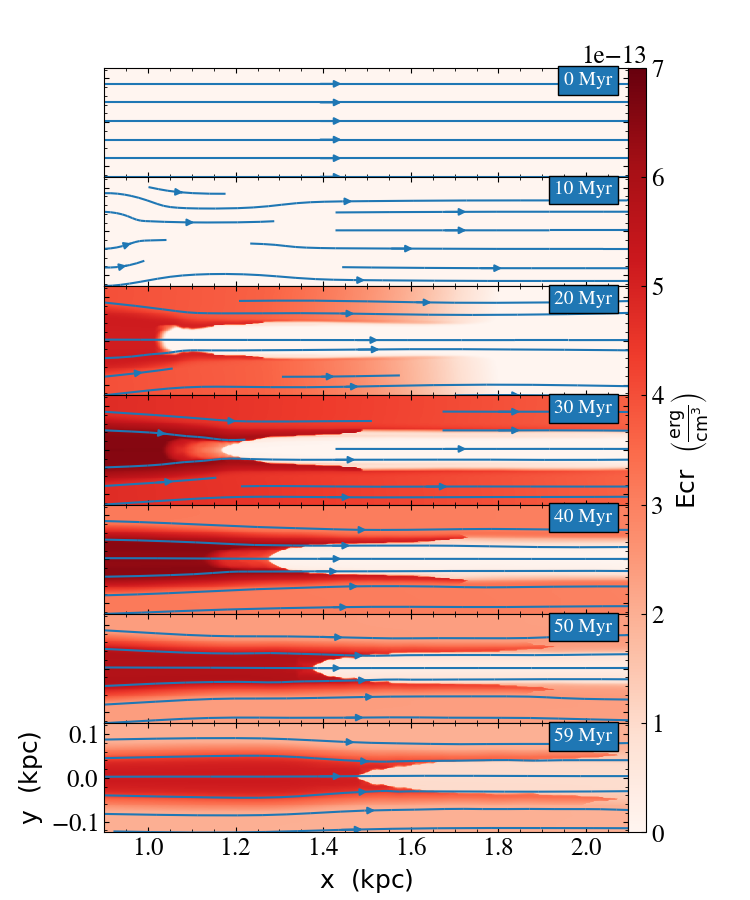}
\caption{Time series slices of the simulation with 3 times the fiducial CR source strength as in Figure \ref{fig:1xseries}. Left: Mass density $\rho$. Right: CR energy density $E_c=3P_c$. Overlaid are streamlines of the magnetic field.}\label{fig:3xseries}
\end{figure*}

\begin{figure*}
\includegraphics[scale=0.35]{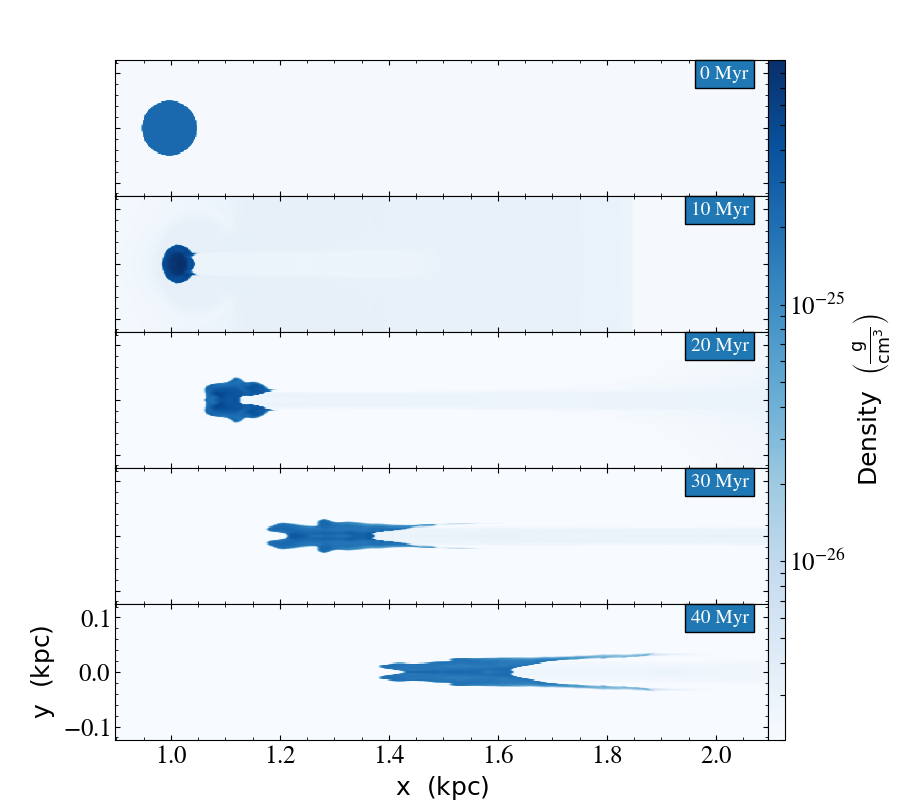}
\includegraphics[scale=0.35]{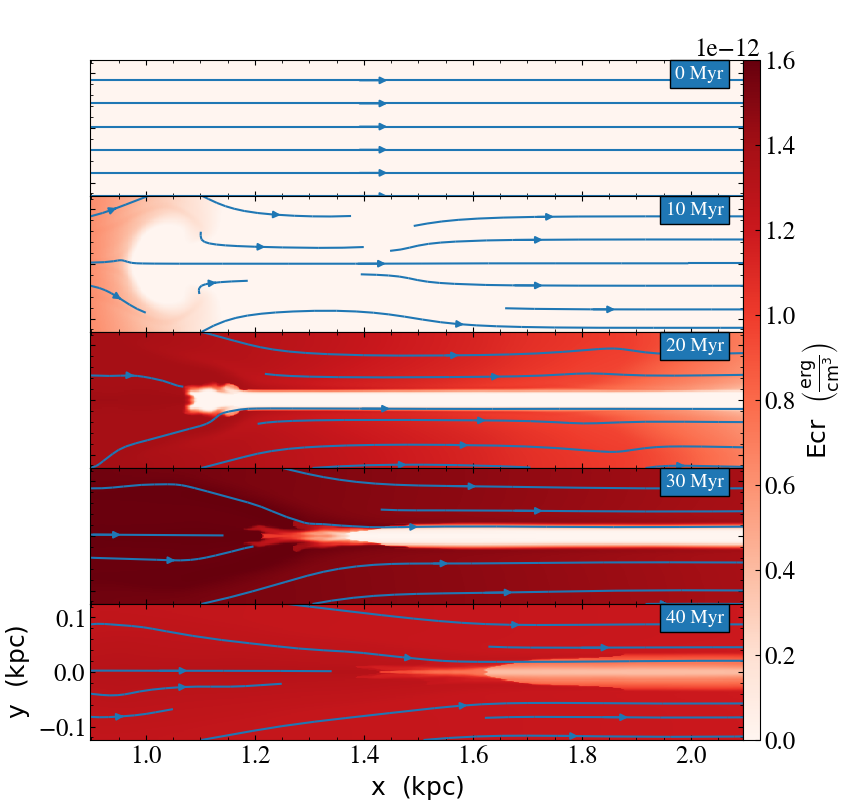}
\caption{Time series slices of the simulation with 10 times the fiducial CR source strength as in Figure \ref{fig:1xseries}. Left: Mass density $\rho$. Right: CR energy density $E_c=3P_c$. Overlaid are streamlines of the magnetic field.}\label{fig:10xseries}
\end{figure*}

We ran similar simulations with identical initial conditions, but with the rate of CR energy injection increased by factors of 3 and 10, in an attempt to accelerate the simulated cloud to higher velocities. Due to computational limitations these were not run for as long in simulated time, but instead were run to 59 and 40 Myr respectively. Time series of density and CR energy with magnetic field streamlines for each are shown in figures \ref{fig:3xseries} and \ref{fig:10xseries}.

\begin{figure}
\includegraphics[width=0.5\textwidth, trim=0cm 1cm 0cm 0cm]{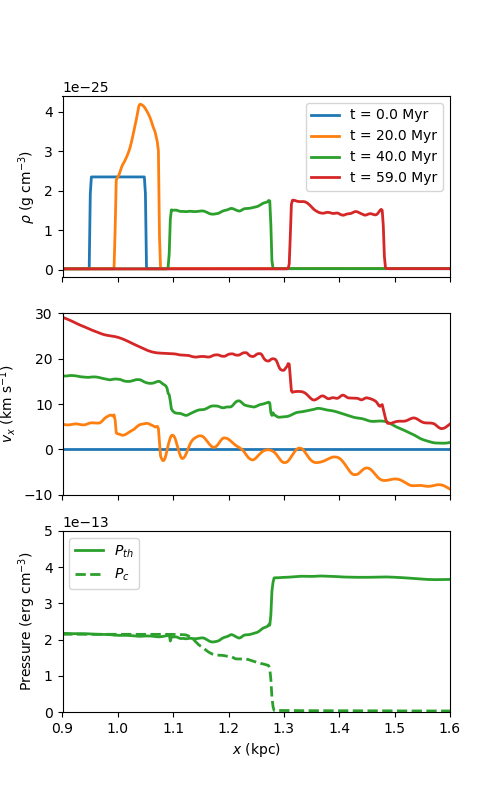}
\caption{Profiles through the center of the cloud ($y=0$) of the simulation with 3 times the fiducial CR source strength. Top: Mass density $\rho$ at various times. Middle: Horizontal velocity $v_x$ at the same times. Bottom: Thermal pressure $P_{th}$ and CR pressure $P_c$ at $t=40$ Myr, when the bottleneck is at its steepest.}\label{fig:3xvpres}
\end{figure}

\begin{figure}
\includegraphics[width=0.5\textwidth, trim=0cm 1cm 0cm 0cm]{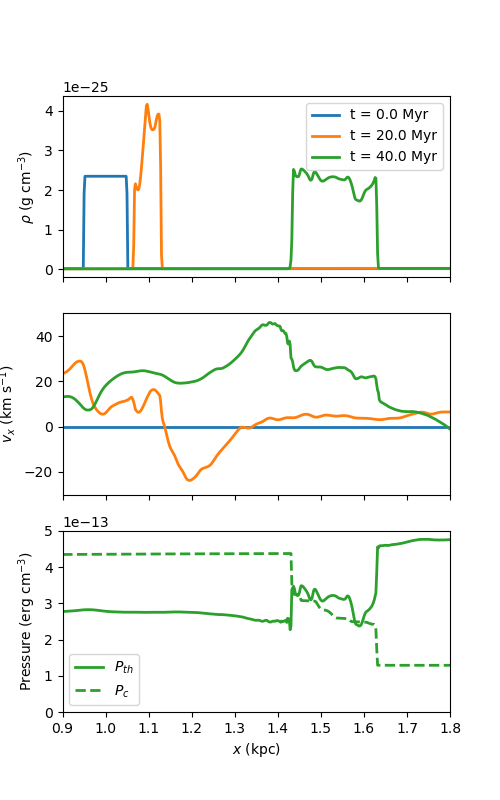}
\caption{Profiles through the center of the cloud ($y=0$) of the simulation with 10 times the fiducial CR source strength. Top: Mass density $\rho$ at various times. Middle: Horizontal velocity $v_x$ at the same times. Bottom: Thermal pressure $P_{th}$ and CR pressure $P_c$ at $t=40$ Myr, when the bottleneck is at its steepest.}\label{fig:10xvpres}
\end{figure}

\begin{figure}
\includegraphics[width=0.5\textwidth]{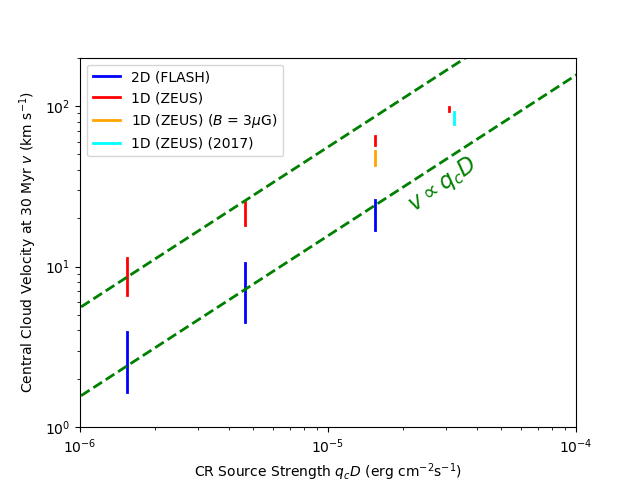}
\caption{Cloud velocities at 30 Myr for different CR source strengths. Each line shows the range of velocities for $T<2\ee{4}$ K gas for one simulation. 2D {\small FLASH} simulations are shown in blue, 1D {\small ZEUS} simulations are shown in red. The 1D run from \protect\citetalias{wiener17a} is shown in cyan, and a 1D run with a 3 $\mu$G field is shown in orange. The green dotted lines show a linear relationship. The cloud velocity increases almost linearly with CR source strength, although falls off from this relation somewhat at large CR source.}\label{fig:qv}
\end{figure}

Qualitatively, these runs are similar to the fiducial run. In all cases the onset of the CR source sends a pressure wave through the domain, followed by the CRs themselves. Both of these push on the cloud, accelerating it to some top speed. At these higher source strengths the cloud is accelerated to higher speeds (about 11 km s$^{-1}$ and 25 km s$^{-1}$ at the end of each run respectively).

With higher CR pressures the 2D effects start to arise. The initial sound wave causes significant flow around the cloud. This causes the magnetic field to roll around the top and bottom of the cloud in a hairpin shape. However, due to flux freezing, the set of field lines which penetrate the cloud stays the same (neglecting the small amount of evaporation or condensation which may occur). As such, the magnetic field topology does not change, and the resulting CR pressure gradients should be roughly the same as the 1D case except that the 2D runs may take longer to build up since CRs have to travel along the longer, rolled-up field lines to reach the cloud.

We do in fact see that the CR pressure upstream of the bottleneck increases significantly as the CR source is increased, as shown in the bottom panels of figures \ref{fig:3xvpres} and \ref{fig:10xvpres}. The 3x run reaches rough equipartition, while in the 10x run CR pressure exceeds the thermal pressure. We also start to see significant CR pressure downstream of the bottleneck. These qualitatively match the behavior of the 1D steady-state solutions in \citetalias{wiener17a}, as well as the 1D time-dependent simulation. The 1D simulation used a source strength 20 times that of the fiducial 2D run, equal to $3.08\ee{-5}$ erg cm$^{-3}$s$^{-1}$.

Before discussing the cloud speeds reached in our simulations, we take a moment to address a numerical peculiarity that emerged in the 2D simulation with the strongest CR source. Figure \ref{fig:10xseries}, right side, shows the CR energy densities at different times for this run. At 20 Myr and beyond, CR pressure gradients have mostly vanished except at the bottleneck itself. This is in contrast to the other two runs, where lateral CR pressure gradients remain throughout the simulation. At first glance this appears to imply that CRs are somehow flowing from the bottlenecked region to the left of the cloud to the areas above and below the cloud. This would itself imply that the magnetic topology of the cloud has changed, that some field lines which started inside the cloud have left it. But this would imply flux freezing was violated, and having checked the field carefully we do not find this to be the case. The vanishing of lateral pressure gradients instead seems to stem from a numerical effect of our outflow boundary conditions on the top and bottom of the domain. If a bottlenecked field line, that is, a field line which enters the cloud, is bent enough that it leaves the domain, this results in extra CRs being added by streaming into the domain boundary nearby. The details of this non-physical mechanism are laid out in appendix \ref{sec:boundary}, but basically it should not affect any of the dynamics along bottlenecked field lines, and so the acceleration of the cloud should not be significantly affected.

The higher pressures are accompanied by faster cloud speeds, shown in the middle panels. Comparing speeds at the same simulation time of 40 Myr between all three simulations, we see that the increase is sub-linear. For CR sources in the ratio 1:3:10, we reach cloud speeds at 40 Myr of around 4, 9, and 25 km s$^{-1}$ respectively. These ratios are quite close to the ratios of maximum cosmic ray pressures achieved between cloud and source. This is further evidence that the transverse spreading that occurs in 2D deprives the cosmic ray piston of some of its force. In addition, a certain amount of magnetic field reconnection takes place due to numerical diffusion. This reduces the cosmic ray flux into the cloud, and hence the cosmic ray momentum available for cloud acceleration. While it is unlikely that numerical diffusion faithfully replicates reconnection physics, it does highlight the sensitivity of the cloud - cosmic ray interaction to magnetic field topology.

Although the 1D runs of \citetalias{wiener17a} are quite different from these 2D runs in many ways, we can put them all on a plot of CR source strength versus cloud velocity. For ease of comparison, we choose a time of 30 Myr to measure the cloud velocity - this is the latest time shown in the 1D run, and the CR time-dependences of the CR sources differ strongly between the 1D and 2D runs after this time (the 1D runs have a steady source throughout rather than the pulsed source described in section \ref{sec:setup}). For each of the 2D runs we take a horizontal line through the cloud center $y=0$ at 30 Myr, and take the minimum and maximum velocities in the cloud, defined as anywhere with temperature less than $2\ee{4}$ K (the resulting velocity range is not sensitive to this cutoff). These velocity ranges are plotted against injected CR flux in blue in figure \ref{fig:qv}. The 1D run in \citetalias{wiener17a} used a source strength twice that of the strongest source used in the 2D runs, equal to $3.08\ee{-5}$ erg cm$^{-3}$s$^{-1}$, and resulted in a cloud speed of about 90 km s$^{-1}$ after 30 Myr. This is shown in cyan, shifted to the right slightly to avoid overlap with other points. We emphasize that spatial resolution, initial cloud position, cloud edge width, magnetic field strength and other properties for this particular run differ from the 2D runs.

For a more apples-to-apples comparison we have also performed new 1D simulations in {\small ZEUS} which more closely reproduce the spatial resolution and initial conditions of the 2D {\small FLASH} runs. These runs use the same initial density and temperature profiles, the same 1 $\mu$G magnetic field, the same time-dependent CR source, and a spatial resolution of 2 pc. However, as with the 1D run from \citetalias{wiener17a}, these runs do not include radiative cooling and instead throw away all energy from CR wave heating. The velocity ranges for these are shown as the red lines in figure \ref{fig:qv}. We have also run one 1D simulation with a higher magnetic field of 3 $\mu$G, shown in orange. A higher magnetic field results in a lower CR pressure for the same injected CR flux (since the Alfv\'en speed at the bottleneck is higher). However, we see that increasing the field by a factor of 3 decreases the cloud velocity by only a little. This suggests that the injected CR flux is more important than CR pressure gradients in determining cloud velocities.

We see a clear upward trend of cloud velocity with CR source strength for both the 1D and 2D runs. The velocity range centers for the 2D runs are about 2.5, 7, and 20 km s$^{-1}$ respectively, which are very close to the 1:3:10 ratio of the CR source strengths, consistent with a linear relationship. This result is promising, and suggests that higher velocities could be reached with proportionally higher CR sources. This conclusion is partially born out by the results of the 1D simulations, which extend to slightly higher CR fluxes. In the 1D runs, we see that the cloud velocity trends linearly with CR source at low values, but falls off somewhat at higher values. The trend among the 2D runs is linear throughout (though there are hints of sublinearity at high CR sources), and all reach lower velocities than their 1D counterparts. This may partially be because radiative cooling increases the mass of the cloud with time relative to the 1D runs (see figure \ref{fig:masscool} and discussion in \S\ref{sec:cool}). Since these mass differences are small, the 2D geometry itself seems to be the likelier cause. Although flux freezing implies that `strength' of the CR bottleneck is unaffected by lateral motions (see discussion in \S\ref{sec:higher}), it is possible that lateral motions reduce the effective ram pressure seen by the cloud from the accelerated ambient medium. In 1D, all rightward gas flow impacts the cloud, but in 2D some is redirected to its sides.

The response of the cloud density to varying CR source strength is far less clear. The three times higher CR source run has a central density profile which drops significantly as the cloud stretches out horizontally, as shown in the top panel of figure \ref{fig:3xvpres}. The magnitude of this effect lies somewhere between that of the fiducial 2D run in the top panel of figure \ref{fig:1xvpres} and the 1D simulation with high CR source in \citetalias{wiener17a}, in which the cloud density drops by as much as a factor of 10. However, in 2D increasing the CR source does not further decrease the cloud density. The density profiles for the ten times higher CR source are shown in the top panel of \ref{fig:10xvpres}, and we can see the central cloud densities at 40 Myr have not significantly decreased from their initial values. 2D effects are likely involved, as the shape of the cloud is quite different in figures \ref{fig:3xseries} and \ref{fig:10xseries}. The proper inclusion of radiative cooling, as opposed to the strategy of throwing away all CR wave heating used in the 1D simulations, is likely a major factor as well.

There are further numerical limitations in these simulations in addition to those discussed in \S\ref{sec:fid}. These runs do not extend as long as the fiducial run in simulation time. This was due to limited computing resources, but is compounded by the fact that at higher CR source strengths, the cloud moves further in the same amount of time, requiring a greater amount of the domain to be at maximum resolution. The extent of the domain in the vertical direction is also limited. In the fiducial run this was of little consequence since the evolution was almost one-dimensional, but at higher CR source strengths lateral motions are significant and we would ideally want to put the lateral domain boundaries farther away.

\subsection{The Role of Radiative Cooling}\label{sec:cool}
In \citetalias{wiener17a} we found that without any radiative cooling to offest the CR wave heating at the bottleneck, clouds can be eroded on the scales of 10s of Myr. We speculated that radiative cooling would efficiently get rid of the heat deposited by the CRs in this way and simply removed this energy in subsequent 1D runs. One of the goals of this work is to determine whether a proper treatment of radiative cooling supports this assumption. The fact that the cloud's edges stay sharp\footnote{As mentioned in \S\ref{sec:fid}, the expected equilibrium length scales are at the pc or sub-pc level and are not completely resolved in these simulations.} as seen in figures \ref{fig:1xseries}-\ref{fig:10xseries} implies that radiative cooling is sufficient to prevent cloud destruction by CR heating at bottlenecks.

\begin{figure*}
\includegraphics[width=0.45\textwidth]{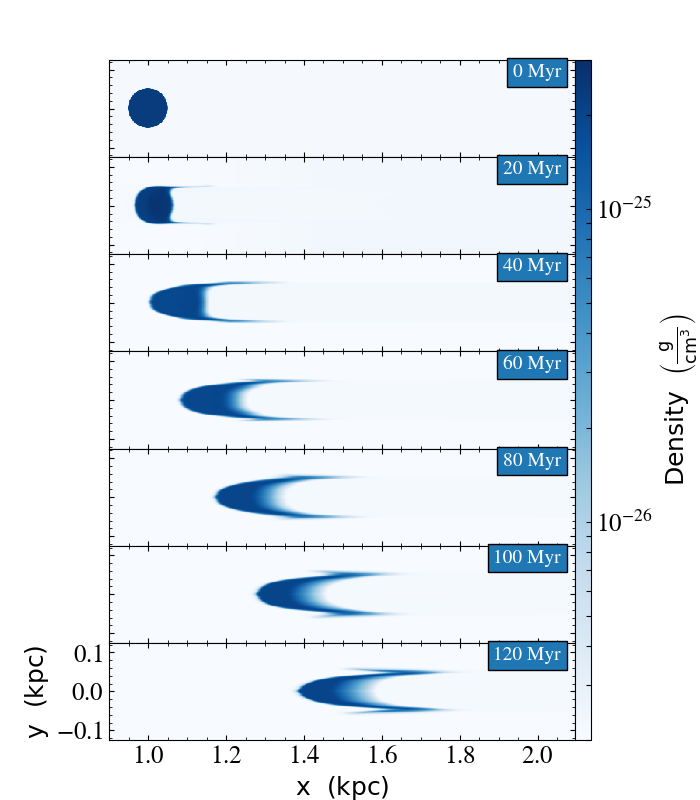}
\includegraphics[width=0.45\textwidth]{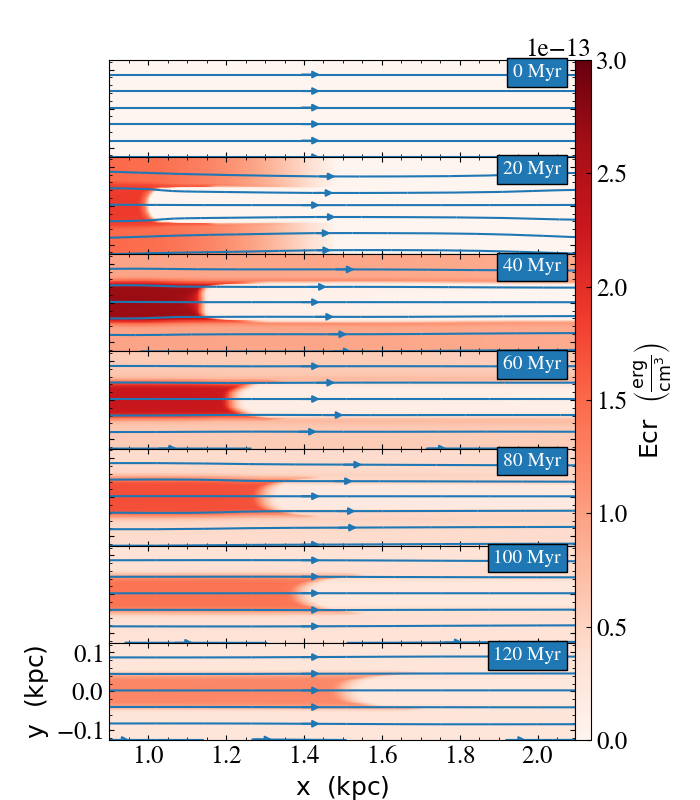}
\caption{Density and CR energy density slices for the run with no radiative cooling. Compare with figure \ref{fig:1xseries}.}\label{fig:ncseries}
\end{figure*}

\begin{figure}
\includegraphics[width=0.5\textwidth]{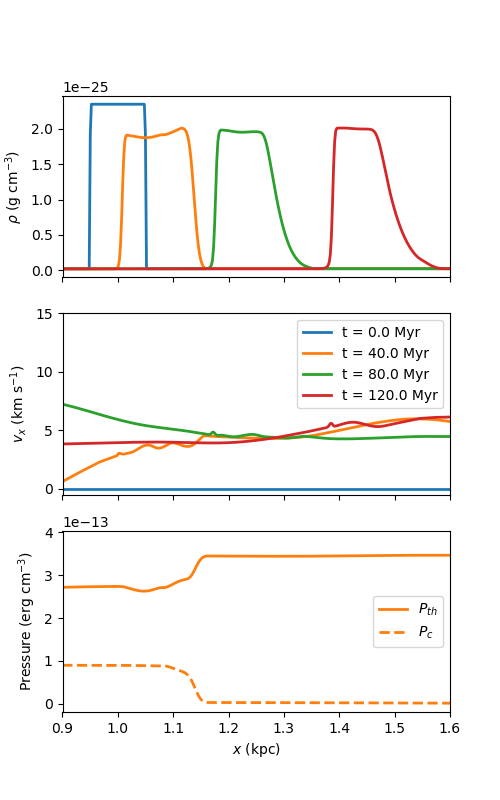}
\caption{Profiles of gas density, horizontal velocity, and thermal and CR pressure along a horizontal line through the center of the cloud for the run with no radiative cooling. Compare with \ref{fig:1xvpres}.}\label{fig:ncpanels}
\end{figure}

\begin{figure}
\includegraphics[width=0.5\textwidth]{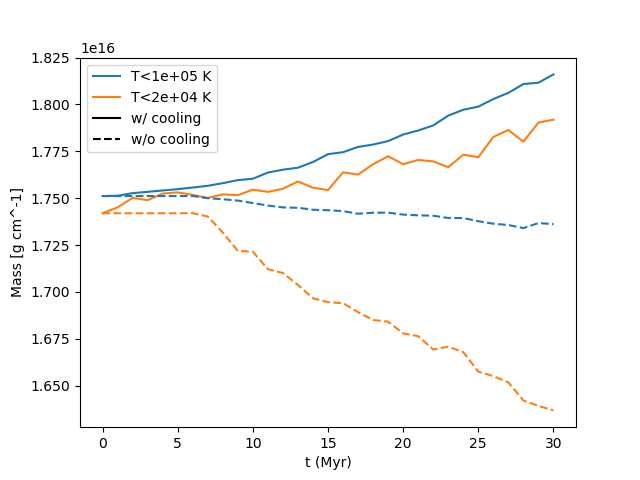}
\caption{Total mass of the cloud as a function of time, defined by two different temperature cutoffs. The solid lines are the fiducial run with radiative cooling, the dashed lines are the run without cooling. The cloud with cooling gains mass due to condensation, while the cloud without cooling loses mass due to CR heating.}\label{fig:masscool}
\end{figure}

To isolate the role that radiative cooling plays we also ran the fiducial simulation with radiative cooling turned off. The results are shown in figures \ref{fig:ncseries} and \ref{fig:ncpanels}. Comparing with the equivalent results from the fiducial run in figures \ref{fig:1xseries} and \ref{fig:1xvpres}, we see the response of the cloud to the onset of CRs is similar. However, without radiative cooling the cloud does not maintain a sharp boundary, and slowly erodes due to CR heating. As a result the total cloud mass drops slightly with time, in contrast to the fiducial cloud which gains mass due to condensation (see figure \ref{fig:masscool}). This reinforces the idea that radiative cooling can keep clouds intact even when accelerated by CR pressure gradients, and is one of the significant conclusions of this work.

The cloud without cooling reaches velocities of 2.1 - 4.6 km s$^{-1}$ in 30 Myr, roughly the same values as the fiducial cloud but slightly higher. This may simply be due to the slightly lower mass of the cloud as it evolves. Peculiarly, unlike the fiducial cloud, the velocity profiles show little spatial variation. While some of this may be attributed to the lack of sharp density gradients, there is little velocity difference between the inside and outside of the cloud, in stark contrast with the fiducial run (see figure \ref{fig:1xvpres} middle panel).

\subsection{Equivalent Star Formation Rates}\label{sec:sfr}
We have discussed the trend of cloud velocity with increasing CR flux, but what CR fluxes might we expect in a Milky Way-like galaxy, or in a starburst galaxy and how do they compare to the values used in this work? To find out, let us roughly convert the Milky Way's star formation rate (SFR) to a CR flux near the disk. The basic picture is as follows. As stars form in the Galactic disk, some fraction of them explode as core-collapse supernovae. Each supernova injects a roughly fixed amount of energy in CRs into the disk. These CRs then flow away from the disk in something approaching plane-parallel symmetry.

Denote the Galactic SFR by $SFR_*$, known to be a few $M_\odot$ yr$^{-1}$, and denote the core-collapse supernova rate per unit stars formed as $R_\tr{SN}$. For typical initial mass functions (IMF) of star formation, we expect approximately 1 core-collapse supernova for every 100-200 $M_\odot$ of stars formed. Each supernova releases about 10$^{51}$ erg of kinetic energy, some fraction $\xi$ of which goes into CRs. Observations of the CR content in the Milky Way suggest $\xi$ is around 0.1. The total injection rate of energy in CRs throughout the Galaxy $I_{cr}$ is then
\begin{multline}
I_{cr}=SFR_*R_\tr{SN}(\xi 10^{51}\tr{ erg SN}^{-1})\\
= \left(\frac{SFR_*}{1\ M_\odot\tr{ yr}^{-1}}\right)\left(\frac{R_\tr{SN}}{1 \tr{ SN}/200 M_\odot}\right)5\ee{47}\tr{ erg yr}^{-1}
\end{multline}
Let us make the gross simplification that this energy is injected uniformly throughout the Galactic disk, which is itself approximated as a thin disk with radius $r_*$ of about 10 kpc. The resulting CR energy flux $\Phi_{cr}$ leaving the disk on both sides is then
\begin{multline}\label{eq:crflux}
\Phi_{cr}=\frac{I_{cr}}{2\pi r_*^2}=3\ee{-6}\tr{ erg cm}^{-2}\tr{s}^{-1}\\ \times\left(\frac{SFR_*}{1\ M_\odot \tr{yr}^{-1}}\right)\left(\frac{R_\tr{SN}}{1 \tr{ SN}/200 M_\odot}\right)\left(\frac{r_*}{10\tr{ kpc}}\right)^{-2}
\end{multline}

This CR flux, which represents the relatively low SFR of the Milky Way, lies in the lower-middle of the range of values tested in these simulations (see figure \ref{fig:qv}). This suggests that significantly higher SFRs such as those in starburst galaxies could drive clouds of the same mass to much faster speeds. M82 for instance has a higher SFR of about 10 $M_\odot$ yr$^{-1}$ (see \cite{gao04}), contained in a much smaller region of radius 200 pc (\cite{forster03}). Converting this to a CR flux according to \eqref{eq:crflux} and extrapolating figure \ref{fig:qv} to this value, we may expect CRs expelled from the starburst region of M82 to accelerate clouds to speeds of ~10$^4$-10$^5$ km s$^{-1}$. This is an optimistic value as it is under the assumption that the CRs drive winds perpendicular to the disk and don't spread out, and also ignores the possibility of clouds being destroyed by flow instabilities. Furthermore, at the higher densities in starburst galaxies, hadronic losses of CRs, which are ignored in this work, would become non-negligible. Nevertheless, equation \eqref{eq:crflux} and figure \ref{fig:qv} suggest that even galaxies with modest SFRs could inject enough CRs to drive a 10$^4$ K cloud of this size to speeds of ~100 km s$^{-1}$.

\subsection{Other Discussion}\label{sec:other}
Figure \ref{fig:qv} neatly summarizes the quantitative result of this work - CR bottlenecks can accelerate clouds to speeds which are roughly proportional to the injected CR flux. But there are clearly other factors which determine the cloud's speed at any given time. We briefly discuss some of these below.

The magnetic field in the vicinity of the cloud will set much of what occurs in its evolution in these simulations. With all other quantities fixed, a higher magnetic field results in a higher streaming speed at the bottleneck. This would lower the steady-state CR pressure for a given CR injection rate, and may result in a slower cloud. A full study of the effects of magnetic field strength would require another set of simulations and is beyond the scope of this paper. However we do have some points of reference for the 1D simulations: the main simulations for comparison to the 2D simulations, which have a 1 $\mu$G field; an equivalent 1D simulation with a 3 $\mu$G field; and the old 1D simulation from \citetalias{wiener17a} which has a 3.26 $\mu$G field (as well as a number of other differences).

We see from figure \ref{fig:qv} that the two runs with higher magnetic field do indeed lie somewhat lower on the plot, i.e. their clouds reach slower speeds than the fiducial 1 $\mu$G field runs. However, the reduction in speed is small, much less than the factor of 3 difference in bottleneck streaming speeds. This suggests that cloud velocity is set more by the incoming CR flux than by the steady-state CR pressure at the bottleneck. On the other hand, with different streaming speeds the timing of events is different, and it may be incorrect to compare runs with different magnetic fields at the same time slices.

Another significant factor for which we have only 1 data point is the total cloud mass. All simulations here have the same column density through the center of the cloud of 7.2$\ee{-5}$ g cm$^{-2}$ (the old 1D run from \citetalias{wiener17a} has a slightly smaller column density of 6.1$\ee{-5}$ g cm$^{-2}$). We therefore cannot make any concrete statements about the dependence of cloud velocity on cloud mass. In multiple dimensions the shape of the cloud is also important, as its cross section determines how wide the CR piston is while its volume is tied to its mass. A flat cloud which is parallel to the disk might reach higher speeds than a spherical cloud of the same mass. We leave the study of these questions to future work.

\subsection{Future Work}\label{sec:future}
There is ample room for extension of these simulations. The greatest computational burden arises from the time constraint required from modeling CR streaming. Using the tanh regularization method of \cite{sharma10}, the simulation time step must be very small,
\begin{equation}\label{eq:sharma}
\Delta t \lsim \frac{\Delta x}{v_s}\frac{\Delta x}{L}
\end{equation}
where $\Delta x$ is the grid cell size, $v_s$ is the CR streaming speed, and $L$ is a smoothing length used in the regularization. The longer $L$ is, the more accurate the simulation, but the harsher the time constraint. For reasonable results we typically require $L$ to be about the size of the domain. For all simulations shown here we use $L=10$ kpc.

However, \cite{jiang18} have recently introduced a new numerical method to handle CR streaming which is based on a radiative transfer method. This method has no smoothing length $L$, and so the associated time constraint is a simple CFL condition
\begin{equation}
\Delta t \lsim \frac{\Delta x}{v_s}
\end{equation}
which is longer than \eqref{eq:sharma} by orders of magnitude. Implementing this method in the future will allow us to address many of the numerical issues present here, such as limited spatial resolution, vertical domain extent, and total simulated time.

\section{Conclusion}\label{sec:conclusion}
We presented a series of two-dimensional MHD simulations exploring the dynamic effects of a CR bottleneck on a cool 10$^4$ K cloud embedded in the hot 10$^6$ K CGM. In earlier 1D simulations (\citetalias{wiener17a}), the CR pressure gradient resulting from the bottleneck was found to accelerate clouds to high velocities ($\sim100$ km s$^{-1}$) and significantly reduce the density inside the cloud. We explored how these results changed with the inclusion of the possibility of lateral motion around the cloud.

We found that at low CR source strengths, lateral pressure gradients are not enough to bend magnetic field lines significantly, and the evolution is largely one-dimensional, and our results resemble those of \citetalias{wiener17a}. We further found that the inclusion of radiative cooling prevents the cloud from being destroyed by CR heating.

At higher CR strengths approaching those required to reach cloud speeds of 100 km s$^{-1}$ in the 1D runs, lateral effects become very important. High pressures induce flow around the cloud, wrapping field lines up around its sides. Despite this, cloud velocity increases for higher CR sources in a nearly-linear correlation up to the source strengths tested here, suggesting it is possible for CR bottlenecks to accelerate clouds to the speeds of observed HVCs. A Milky Way sized galaxy with a star formation rate of 20 - 30 $M_\odot$ yr$^{-1}$ would generate enough CRs to accelerate clouds to 100 km s$^{-1}$. However, limited computational resources make further testing prohibitive until the development of CR bottleneck simulations which take advantage of new numerical methods for streaming.

\section*{Acknowledgements}
We acknowledge Marcus Br\"uggen for useful discussions. We acknowledge Karen Yang for her support with using the {\small FLASH} code. We acknowledge Chad Bustard for additional code support as well as supplying the analytic fit to the cooling curve. Resources supporting this work were provided by the NASA High-End Computing (HEC) Program through the NASA Advanced Supercomputing (NAS) Division at Ames Research Center. JW and EGZ acknowledge support by NSF Grant AST-1616037, the WARF Foundation, and the Vilas Trust. MR acknowledges support by NSF Grant AST-1715140.

\bibliographystyle{mnras}
\bibliography{master_references2}

\appendix
\section{Boundary-Related Issues}\label{sec:boundary}
In \S\ref{sec:higher} we alluded to a numerical issue with our boundary conditions which resulted in the erasure of lateral CR pressure gradients in our last 2D run shown in figure \ref{fig:10xseries}. We describe here the details of this issue, which occurs wherever there is a lateral CR pressure gradient across adjacent field lines which intersect the top or bottom domain boundary at an angle other than 90 degrees.

\begin{figure}
\includegraphics[width=0.5\textwidth]{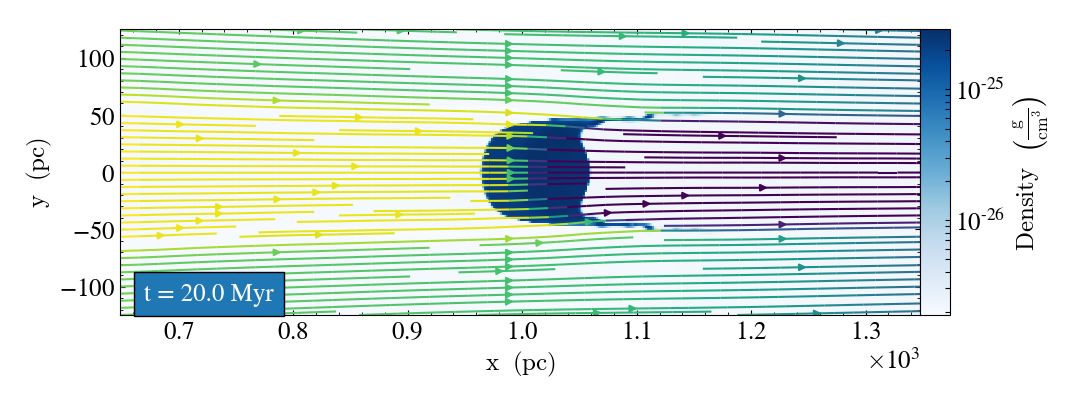}
\includegraphics[width=0.5\textwidth]{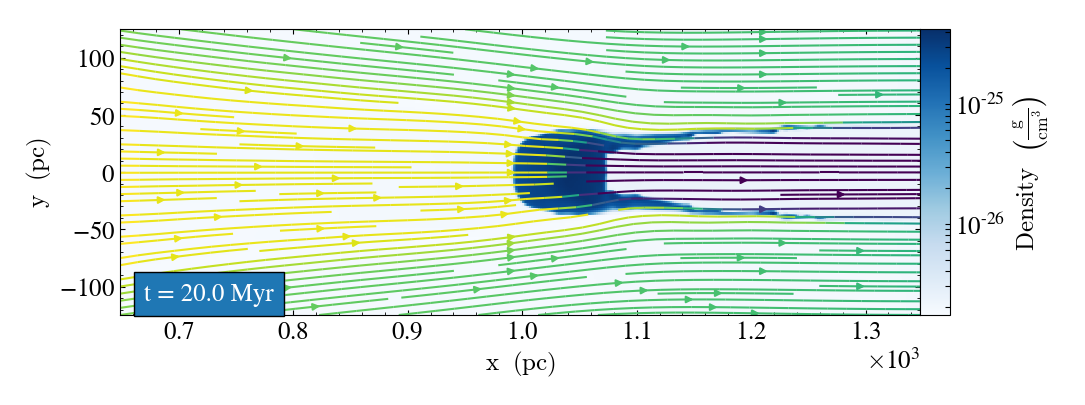}
\includegraphics[width=0.5\textwidth]{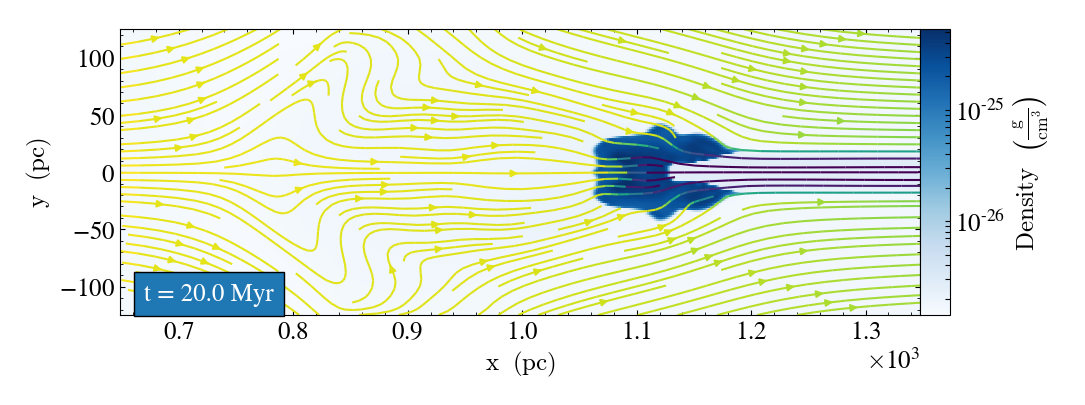}
\caption{Plots of density at 20 Myr for each of the three 2D runs, overlaid with magnetic field streamlines. The lines are colored according to CR energy density, such that a color contrast indicates a CR pressure gradient. A vertical CR gradient is seen to the left of the cloud for the top two panels but not for the bottom panel.}\label{fig:lines}
\end{figure}

First we present evidence that field lines leaving the domain are the cause of the loss of lateral CR pressure gradients seen in \ref{fig:10xseries}. Figure \ref{fig:lines} shows a gas density plot at 20 Myr for each of the three 2D runs presented in this work. Overlaid on top of the density are streamlines which follow the magnetic field and are colored according to CR pressure. Yellow implies high CR pressure, while purple indicates low CR pressure.

In the first two runs, the lateral CR pressure gradients clearly remain, as seen by the color contrast between field lines. As seen from the plots, the border between the high CR pressure in yellow and the lower CR pressure in green is formed by the top- and bottom-most magnetic field lines which still intersect the cloud. This is consistent with the idea that CRs streaming along field lines which intersect the cloud are effectively bottlenecked, while those streaming along other field lines can stream faster. Note that none of the bottlenecked field lines leave the domain out the top or bottom.

In the third run, however, there is no lateral color contrast seen, indicating there are no CR pressure gradients perpendicular to the field lines. Note that the field lines where we would expect such a gradient to be present, which again are the top- and bottom-most lines which intersect the cloud, also leave the domain out the top and bottom.

\begin{figure*}
\includegraphics[width=0.45\textwidth]{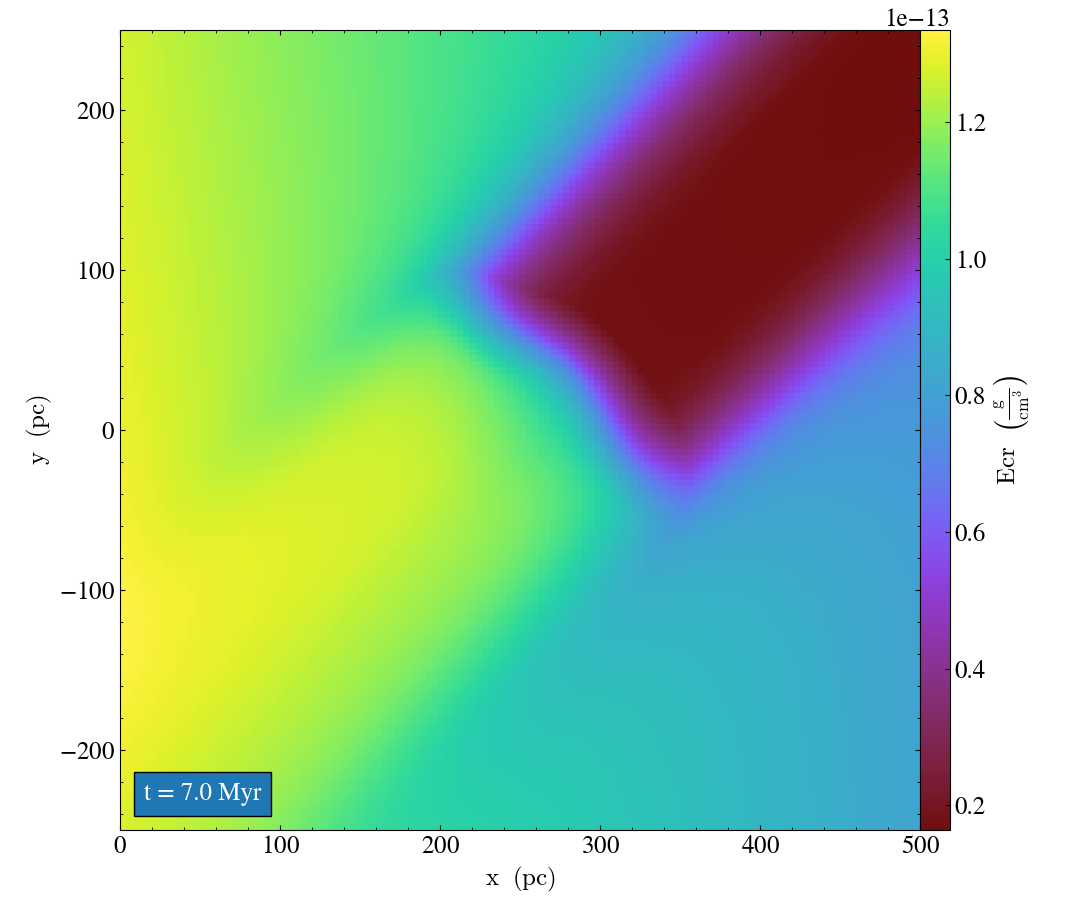}
\includegraphics[width=0.45\textwidth]{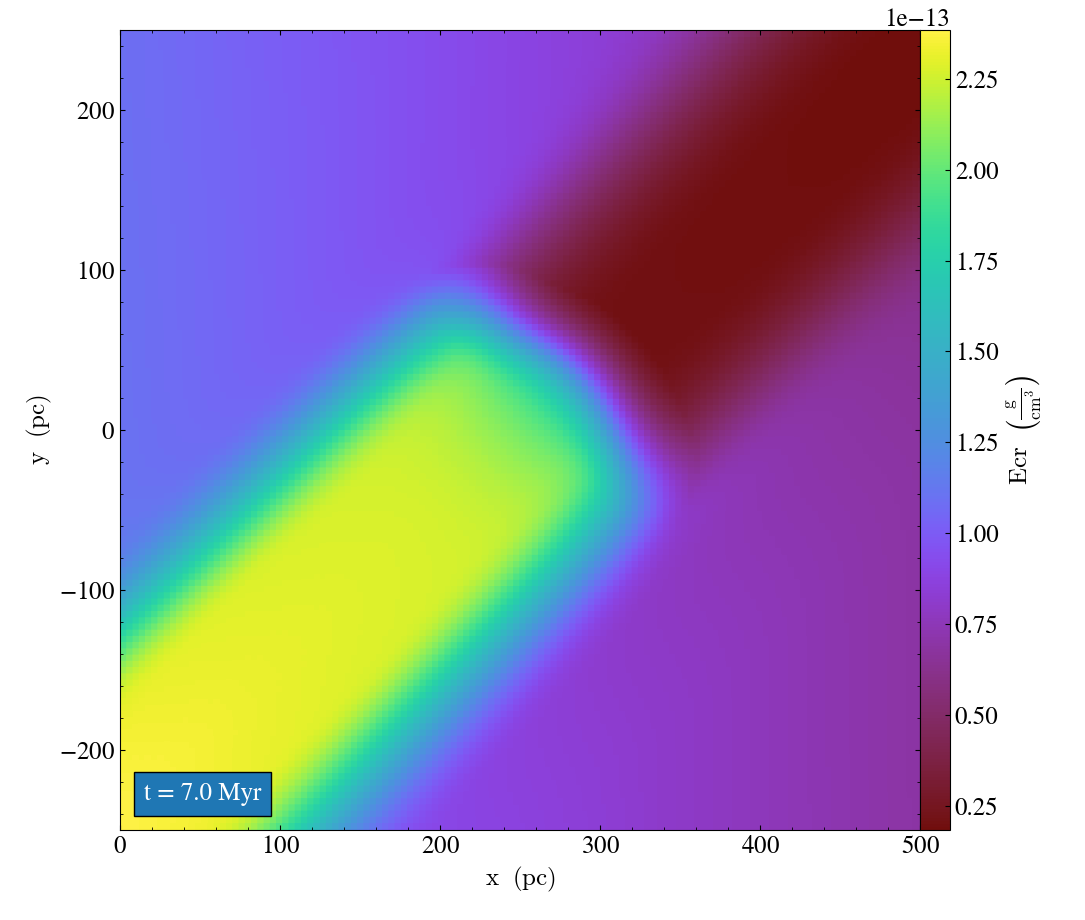}
\caption{An early test of the {\small FLASH} code with the MHD turned off. CRs are injected via a source term into the leftmost active domain cells and stream along tilted magnetic field lines toward a dense cloud in the center of the domain. Left: Result of normal reflecting boundary conditions. Right: Results of reflecting boundary conditions with the CR flux across the left boundary forced to zero.}\label{fig:tilt}
\end{figure*}

This removal of cross-field CR gradients appeared in one of our very early tests of the {\small FLASH} code, where we injected CRs along the left boundary and had them stream down a titled B-field toward a dense cloud. In these tests we turned off the MHD and only evolved the CRs in time. The top and bottom boundaries were periodic, and the left boundary was reflecting. We expected the CRs to build up in pressure on field lines which entered the cloud. Instead we saw roughly the same CR pressure everywhere, except for the reduced pressure past the bottleneck (see figure \ref{fig:tilt}, left). This did not fit with our theory.

\begin{figure*}
\includegraphics[width=0.7\textwidth]{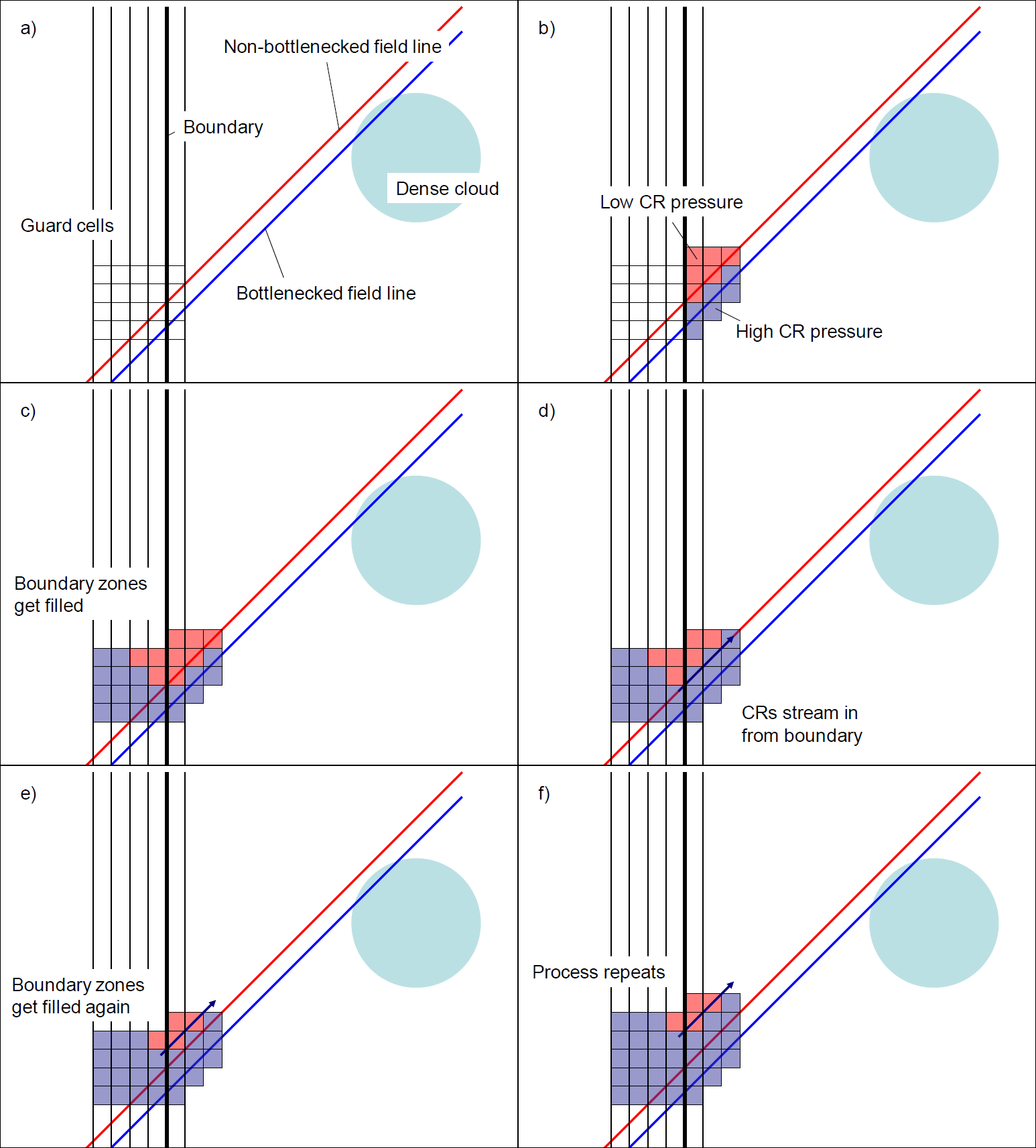}
\caption{Schematic showing how the outflow boundary condition can lead to extra CRs entering the domain via streaming. High CR pressures on a bottlenecked field line are copied into the guard cells, where CRs can stream into the domain along adjacent, non-bottlenecked field lines.}\label{fig:diagram}
\end{figure*}

We found that our results were consistent with our theory only if we negated any CR flux through the left boundary (see figure \ref{fig:tilt}, right). This suggested there was some unintended CR flux through the boundary. A schematic for how the boundary conditions cause this is shown in figure \ref{fig:diagram}. This figure illustrates the issue when reflecting boundary conditions are used, but the same problem arises when using outflow boundary conditions. (a) Consider two adjacent field lines, one which enters the cloud (shown in blue) and one which doesn't (shown in red). (b) CR pressure will build up on one line because of the bottleneck effect, but not on the other. Once this buildup reaches the boundary, the boundary conditions fill in the guard zones (c) according to the condition chosen. Outflow conditions will copy values from the first active cell to all the guard cells. Reflecting conditions will mirror the values of the first active cells, with normal the gas velocity component flipping sign (normal magnetic field is not flipped). Guard cells across the boundary from a bottlenecked field line will be given large CR pressures. But this induces a CR gradient along the adjacent field line in red, causing CRs to flow into the domain along this line via streaming (d). This raises the CR pressure on a field line even though it doesn't pass through the cloud. This unphysically high CR pressure then propagates to further field lines (e,f).

\begin{figure*}
\includegraphics[width=0.7\textwidth]{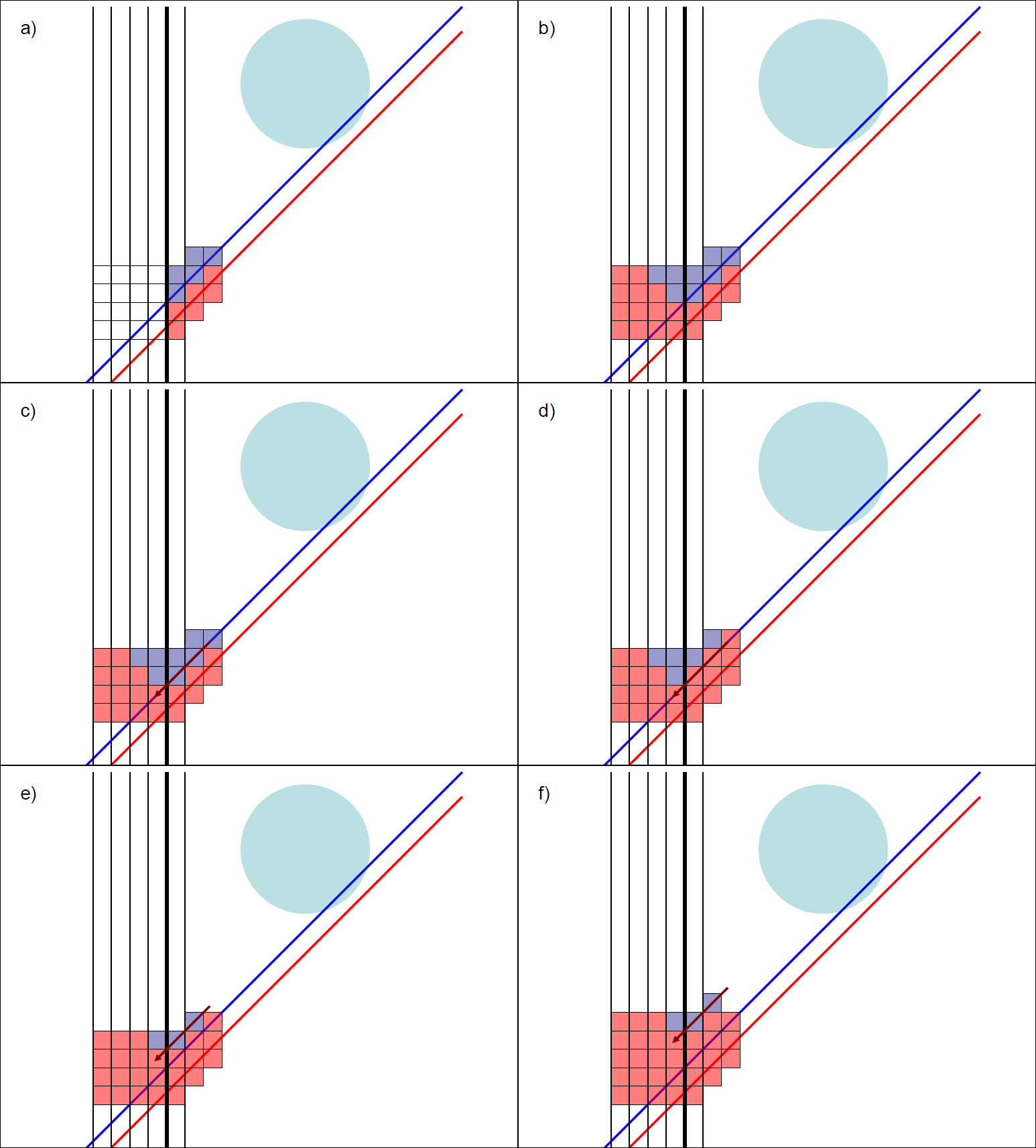}
\caption{Same as figure \ref{fig:diagram}, but showing how CRs can spuriously leave the domain.}\label{fig:diagram2}
\end{figure*}

The same effect can actually cause CRs to leave the domain unphysically as well. This is shown in figure \ref{fig:diagram2}, which considers two adjacent field lines on the opposite end of the cloud. Here the boundary conditions (either outflow or reflecting) introduce a boundary cell with low CR pressure along one of the bottlenecked field lines. The resulting pressure gradient causes CRs to stream out of the domain, reducing the pressure in the bottleneck. This explains why the CR pressure in the bottlenecked region of figure \ref{fig:tilt} left is less than that of figure \ref{fig:tilt} right despite using the same setup.

We applied the fix of negating all CR streaming flux through the left boundary to avoid this problem. We also ended up negating advective flux of CRs through this boundary for a different reason - to control the amount of CRs being added to the domain. However, as we wanted CRs to be able to stream out of the top and bottom of the domain, the streaming flux was not negated there. Unexpectedly, in the third 2D simulation field lines connected to the cloud left the domain through the top and bottom, so this issue reappeared in a different form. This simulation would properly be redone with a domain which is wider in the vertical direction to accommodate the spreading of the field lines.

According to the geometry of the magnetic field lines leaving the domain in our cloud simulations, e.g. figure \ref{fig:lines} bottom panel, we only expect CRs to be spuriously added (as in figure \ref{fig:diagram}) to the domain along non-bottlenecked field lines. We don't expect any CRs to be spuriously removed (as in figure \ref{fig:diagram2}) anywhere. This implies that the CR dynamics along bottlenecked field lines are not significantly impacted by these boundary issues. Hence, we do not expect the acceleration of the cloud to be affected much.

\section{Resolution Test}\label{sec:resolution}
\begin{figure*}
\includegraphics[width=0.45\textwidth]{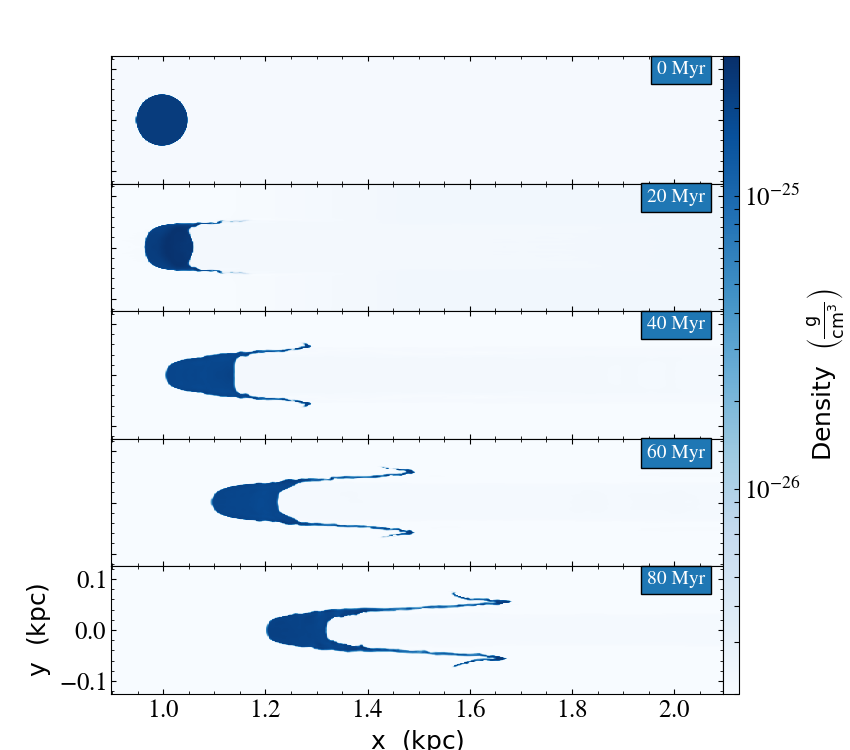}
\includegraphics[width=0.45\textwidth]{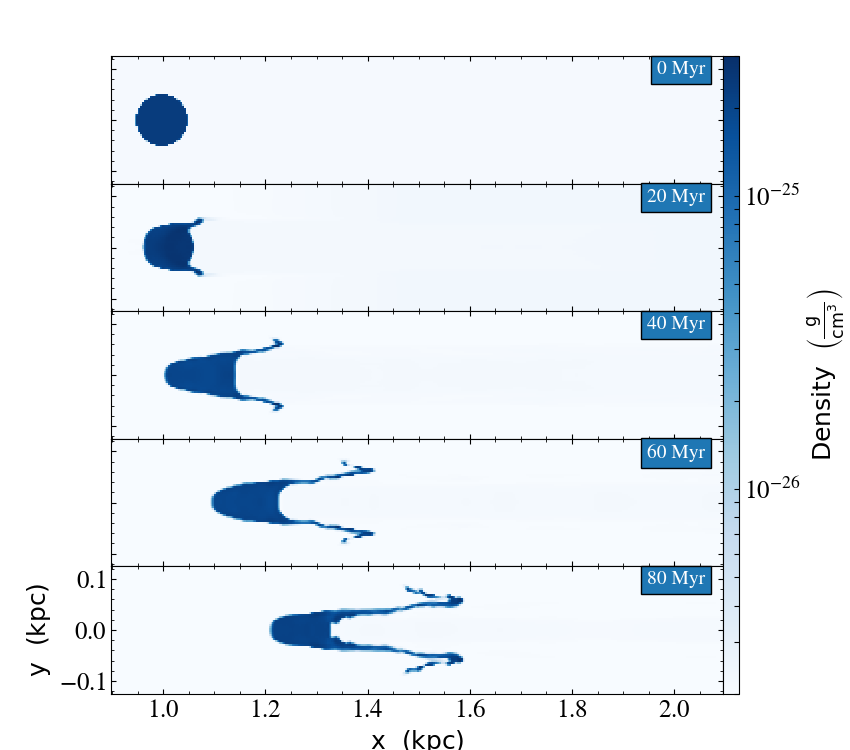}
\caption{Slices of density at different times for simulations at two different spatial resolutions. Left: The fiducial run (also shown in figure \ref{fig:1xseries} left for more times). Right: The same setup with two times lower spatial resolution.}\label{fig:densres}
\end{figure*}

\begin{figure*}
\includegraphics[width=0.45\textwidth]{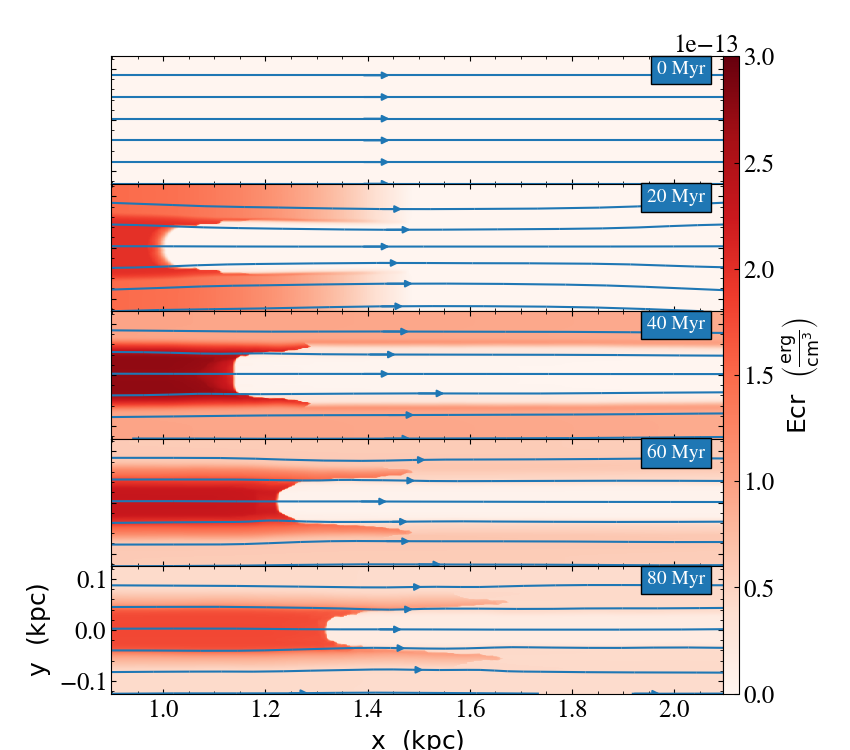}
\includegraphics[width=0.45\textwidth]{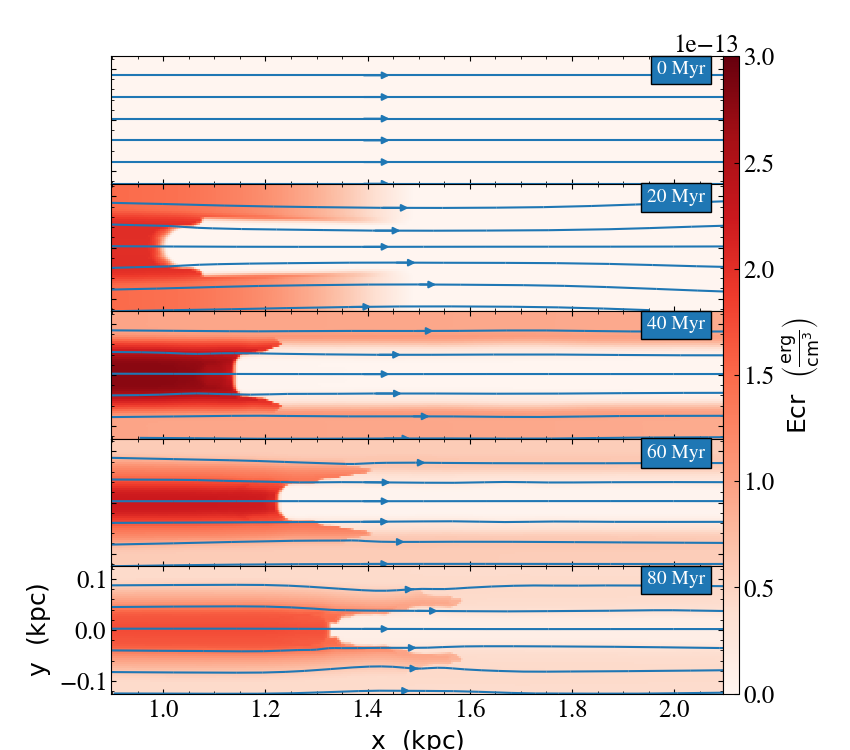}
\caption{Slices of CR energy density at different times for simulations at two different spatial resolutions, with some magnetic field lines overlaid. Left: The fiducial run (also shown in figure \ref{fig:1xseries} right for more times). Right: The same setup with two times lower spatial resolution.}\label{fig:crres}
\end{figure*}

We unfortunately did not have enough computing resources to do a proper resolution test. Instead we present a qualitative test wherein we ran the fiducial run at half the spatial resolution. Density slices for this run are shown in figure \ref{fig:densres} alongside the equivalent time slices of the fiducial run. Similarly, CR energy densities with magnetic field lines overlaid are shown in \ref{fig:crres}. While there are some clear differences, the overall behavior is largely the same between the two resolutions.

\begin{figure}
\includegraphics[width=0.5\textwidth]{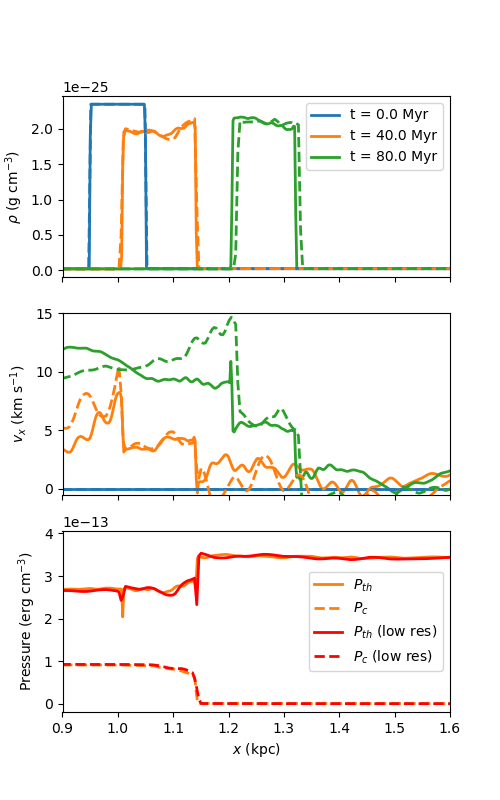}
\caption{Profiles of various quantities along a horizontal line through the center of the cloud at various times. Top: Gas density. The fiducial run is shown in solid lines, the low resolution run in dashed lines. Middle: Velocity in the horizontal direction. The fiducial run is shown in solid lines, the low resolution run in dashed lines. There is some disagreement outside the cloud, but the velocities of the cloud itself are in good agreement. Bottom: Thermal (solid) and CR (dashed) pressures at 40 Myr. The fiducial run is shown in orange and the low resolution run in red.}\label{fig:xvres}
\end{figure}

We can make a more direct comparison by plotting quantities along the horizontal line through the center of the cloud as in figure \ref{fig:1xvpres}. This is shown in figure \ref{fig:xvres}. The top two panels show the density and horizontal velocity profiles respectively. The solid lines show the fiducial run, and the dashed lines show the lower resolution run. There is very good agreement between the density profiles. Agreement is not as good between the velocity profiles, but they are close to each other in the cloud region.

The bottom panel of figure \ref{fig:xvres} shows the thermal pressure profiles (in solid lines) and CR pressure profiles (in dashed lines) at 40 Myr, with the fiducial run in orange and the low resolution run in red. These also show very good agreement.

\end{document}